\documentclass[preprint,11pt]{elsarticle}

\usepackage[english]{babel}

\usepackage[letterpaper,top=2cm,bottom=2cm,left=3cm,right=3cm,marginparwidth=1.75cm]{geometry}

\usepackage{amsmath}
\usepackage{mathrsfs}
\usepackage{amssymb}
\usepackage{bm}
\usepackage{comment}
\usepackage{graphicx}
\usepackage{subcaption}
\usepackage[colorlinks=true, allcolors=blue]{hyperref}
\usepackage{adjustbox}
\usepackage{float}
\usepackage{tikz}
\usetikzlibrary{arrows.meta}
\usepackage{caption}
\usepackage{svg}
\usepackage{soul}
\usepackage{xcolor}
\usepackage{comment}
\usepackage{units}
\usepackage{ulem}
\usepackage{lineno}
\usepackage{tikz}
\usepackage{graphicx} 
\usepackage{subcaption}   
\usepackage{tcolorbox}
\usepackage{minted}
\usepackage{ulem}
\usepackage{cancel}
\usepackage{adjustbox} 
\usetikzlibrary{matrix} 
\newtcbox{\inlinecode}{on line, 
	colback=gray!10,  
	colframe=gray!50, 
	rounded corners,
	boxrule = 0.5pt,
	left = 1pt, right=1pt, top=0.5pt, bottom=0.5pt,
	fontupper = \ttfamily\small,
	boxsep=0pt
	}
\newcommand{\inlinearrow}{%
  \tikz[baseline=-0.5ex] \draw[<->,>=stealth,line width=2pt] (0,0) -- (0.6,0);%
}

\DeclareMathAccent{\svec}{\mathord}{letters}{126}

\journal{arXiv}

\begin{document}

\begin{frontmatter}
\title{Accelerating high-order energy-stable discontinous Galerkin solver using auto-differentiation and neural networks}

\author[1]{Xukun Wang}
\cortext[cor1]{Corresponding author}
\ead{xukun.wang@alumnos.upm.es}
\author[1]{Oscar A. Marino}
\author[1,2]{Esteban Ferrer}

\address[1]{ETSIAE-UPM-School of Aeronautics, Universidad Politécnica de Madrid, Plaza Cardenal Cisneros 3, E-28040 Madrid, Spain}
\address[2]{Center for Computational Simulation, Universidad Politécnica de Madrid, Campus de Montegancedo, Boadilla del Monte, 28660 Madrid, Spain}


\begin{abstract}

High-order Discontinuous Galerkin Spectral Element Methods (DGSEM) provide excellent accuracy for complex flow simulations, but their computational cost increases sharply with higher polynomial orders. 
To alleviate these limitations, this work presents a differentiable DG solver coupled with neural networks (NNs) that learn corrective forcing terms to correct low-order simulations and provide high-order accuracy. The solver's full differentiability enables gradient-based optimization and interactive (solver-in-the-loop) training, mitigating the data-shift problem typically encountered in static, offline learning. Two representative test cases are considered: the one-dimensional viscous Burgers' equation and two-dimensional decaying homogeneous isotropic turbulence (DHIT). The results demonstrate that interactive training with extended unrolling horizons substantially improves the precision and long-term stability of the simulation compared to static training. For the Burgers' equation, a $\mathbb{P}_2$ simulation corrected using a NN-correction achieves the accuracy of a $\mathbb{P}_4$ solution with eight times reduction in computational cost. For the DHIT case, the NN-corrected low-order simulations successfully achieve high-order accuracy while reduce the error beyond the training interval. These results highlight the potential of differentiable solvers combined with neural networks as a robust and efficient framework for accelerating high-fidelity DG-based fluid simulations.

\end{abstract}

\end{frontmatter}

\tableofcontents
\section{Introduction}

The high-order Discontinuous Galerkin method (DG) has emerged as a powerful framework for solving partial differential equations (PDEs), combining the geometric flexibility of finite elements with the high accuracy of spectral methods. By employing piecewise polynomial approximations within elements and enforcing conservation through numerical fluxes, DG methods excel in solving conservative systems in complex geometries \cite{cockburn_tvb_1989,bassi_high-order_1997,toulopoulos_high-order_2006,ferrer_high_2012,FERRER2023108700}. As a specific form of the DG framework, Discontinuous Galerkin Spectral Element Methods (DGSEM) \cite{kopriva2009implementing, karniadakis_spectralhp_2005} combine the flexibility of DG with spectral accuracy, leading to exponential convergence for smooth solutions, allowing precise resolution of multiscale phenomena in computational fluid dynamics (CFD) \cite{beck_highorder_2014}. However, higher polynomial order comes with increased computational costs: high-order quadrature rules lead to additional operation counts, and explicit time-stepping schemes necessitate more restrictive Courant–Friedrichs–Lewy (CFL) conditions. These bottlenecks intensify as the size of the problem increases, underscoring the need for effective speed-up techniques. 

The emergence of the fourth paradigm of scientific discovery \cite{hey_fourth_2012} -- characterized by data-driven modeling and large-scale computational experimentation -- has laid the foundation for machine learning (ML) to revolutionize the way of research in fluid dynamics. ML has been widely used in the field of fluid dynamics\cite{annurev:/content/journals/10.1146/annurev-fluid-010719-060214}, including flow feature extraction\cite{champion_data-driven_2019,lusch_deep_2018}, turbulence modeling\cite{annurev:/content/journals/10.1146/annurev-fluid-010518-040547, Beck_perspective}, super-resolution\cite{fukami_super-resolution_2023}, flow control\cite{novati_controlled_2019,verma_efficient_2018}, and aircraft optimization design \cite{LECLAINCHE2023108354}. In addition, ML techniques have been widely used to improve the performance of traditional numerical solvers to improve accuracy and decrease the simulation cost. For example, Bar-Sinai et al. \cite{bar-sinai_learning_2019} develop a data-driven discretisation of the spatial derivatives in a low-resolution mesh learned from a high-resolution solution by a neural network (NN). Following the same ideal, Kochkov et al.\cite{Kochkove2101784118} accelerate the DNS of the two-dimensional Kolmogorov flow\cite{chandler_invariant_2013} by learning interpolation or correction using a convolutional neural network (CNN) in an incompressible solver based on the finite-volume method (FVM). De Lara and Ferrer \cite{de_lara_accelerating_2022} first introduce the idea of adding NN-modeled correction to the DGSEM framework to accelerate the high-order simulation of the one-dimensional Burgers' equation. This method has been successfully extended to solve the three-dimensional Navier--Stokes equations (NS) in the cases of the Talyor Green vortex (TVG) \cite{manrique_de_lara_accelerating_2023} and the channel flow\cite{marino_comparison_2024,marino_accelerating_2024}. 

Despite the great potential, the widely-used offline correction learning methods are prone to the problem of stability and error accumulation. First, recurrent calls to NN will enlarge the negligible initial error to often unbearable levels, degrading the accuracy of the entire simulation. In addition, the data shift\cite{wiles_fine-grained_2021}, defined as the mismatch between the training and inference data, makes the performance of NN in the inference phase not as good in the training phase. This phenomenon is also reported for NN-modeled turbulence models\cite{duraisamy_perspectives_2021}. 
Several strategies have been explored to overcome these problems. The simplest method generates training data in advance that is closer to the inference scenario, classified as the data enhancement approach, including precomputed interactions\cite{um_solver---loop_2020} and adding noise disturbance to the training data\cite{duraisamy_perspectives_2021, kiener_correcting_2025}. This method improves accuracy, but cannot alleviate the accumulation of errors and has the risk of degrading the learned NN model. In \cite{kiener_correcting_2025}, reinforcement learning (RL) has been investigated as an alternative approach to handle this problem showing potential accuracy improvements but non-negligible increase in cost. The most effective method proposed to date is to use an end-to-end differentiable solver, which makes it possible to back-propagate the gradients throughout the numerical solver during the training phase. The differentiability of solver enables one to minimize the data-shift by enlarging unrolling horizons, leading to a more stable and accurate modeling of the target trajectory. To the best of our knowledge, the differentiable low order solver was first proposed to learn corrections by Um et al. \cite{um_solver---loop_2020}, in the so-called `Solver-in-the-Loop' approach. Similarly, JAX-CFD has also been coupled with NN training in \cite{Kochkove2101784118}. Bezgin et al.\cite{bezgin_jax-fluids_2023} developed the first differentiable solver for multi-phase simulations. Later, it attracted great attention in turbulence modeling \cite{chen_learned_2022, greif_physics-preserving_2023, shankar_differentiable_2024}. Hugo et al. \cite{melchers_comparison_2023} compared three types of model closure strategies in the one-dimensional Burger's equation and the Kuramoto-Sivashinsky equation and found that `“trajectory fitting” with discretization', followed by optimization, is the best choice in terms of accuracy. More recently, List et al. \cite{list_differentiability_2025} proposed a remedy to incorporate the gradients across time steps into a non-differentiable solver, but pointed out that interactive correction learning coupled with a differentiable solver is still the optimal approach. 

In this paper, we develop a differentiable high-order DGSEM solver and accelerate the high-order DG simulation by coupling it with the corrective forcing learned by NN as proposed by de Lara and Ferrer \cite{de_lara_accelerating_2022}. The potential of this combination is explored for the one-dimensional Burgers' equation and the two-dimensional decaying homogeneous isotropic turbulence (DHIT). The propagation of gradients of static and interactive training is analyzed in detail. To avoid widely observed biases \cite{mcgreivy_weak_2024} in the field of ML-fluid cross-research, we fairly compare the error evolution between simulations corrected for NN and normal simulations using different polynomial orders. The accuracy of the NN-corrected simulation is rigorously measured. The results still show the great potential of this approach in reducing the low-order simulation error. 

The remainder of the paper is organized as follows. Section 2 details the proposed methodology, including a brief introduction to the numerical scheme and the NN training methods. Both, static and interactive training strategies are introduced and analyzed. Section 3 provides the simulation results, including the one-dimensional Burgers' equation and the two-dimensional DHIT. Finally, the conclusion and future outlook are given in Section 4.

\section{Methodology}
\subsection{Discontinuous Galerkin Spectral Element Method}\label{sec:dgsem}

We discretise the equations (Burgers, Navier-Stokes) using the Discontinuous Galerkin Spectral Element Method (DGSEM), which is a particularly efficient nodal version of DG schemes \cite{kopriva2009implementing, FERRER2023108700}. In addition to enhance stability, we use Pirozzoli's energy stable formulation \cite{pirozzoli_numerical_2011}. Since the code is written in \inlinecode{JAX}, the resulting solver is auto-differentiable.
For simplicity, let us consider a one-dimensional advection-diffusion conservative law:
\begin{align}
    \label{eq: 1d conservation law}
     \frac{\partial q}{\partial t} + \frac{\partial f_e}{\partial x} &= \frac{\partial f_v}{\partial x},\; q\in \Omega\times[0,T] \\
     q(x;0) &= q_0(x) \\ 
     q(x;t) &= q_b(x;t),\; x\in \partial\Omega, t\in [0,T],
\end{align}
where $\Omega\subset \mathbb{R}$ is the spatial physical domain, $x \in \Omega$ are the spatial coordinates, $t \in [0,T]$ is the time, $q(x,t): \Omega\times[0,T]\to \mathbb{R}^s$, where $s$ is the number of variable components in conservative variables $q$. $f_e$ and $f_v$ are the inviscid and viscous fluxes, respectively, and $q_0(x)$ and $q_b(x;t)$ are the initial and boundary conditions, respectively. The solution $q$ is approximated in the space:
\begin{equation}
    V^p_h = \{v \in L^2({\Omega})\; \big\vert\; v\vert_{\Omega_i} \in \mathbb{P}_p(\Omega_i),\; \forall\; \Omega_i \in \mathcal{T}_h\},
\end{equation}
where $\mathcal{T}_h$ is a tessellation  of the domain $\Omega$ into non-overlapping $K$ elements $ \Omega_i = [x_{i-1},x_i],i=1,\dots,K$ (mesh); $\mathbb{P}_p(\Omega_i)$ is the space of polynomials of degree $\leq p$ on $\Omega_i$, and $x_i, i =0,\dots,K$ denote the location of faces of elements.

In practice, DGSEM solves the equation in a reference element, $\Omega_{\text{ref}} = [-1,1]$, which is geometrically transformed from a physical element $\Omega_{i}= [x_{i-1},x_i]$ through a transfinite mapping:
\begin{equation}
    \xi = \xi_i(x) = \frac{2}{\Delta x_i}\left( x - \frac{x_{i-1}-x_{i}}{2}\right),
\end{equation}
where $\Delta x_i = x_i - x_{i-1}$ is the length of the element $\Omega_{i}$. The transformation is applied to Eq.~\eqref{eq: 1d conservation law}, with the result that the following:
\begin{equation}
    \label{eq:1dConservationTransformed}
    \mathcal{J}\frac{\partial q}{\partial t} + \frac{\partial f_e}{\partial \xi} = \frac{\partial f_v}{\partial \xi},
\end{equation}
where $\mathcal{J} = \partial x/\partial \xi $ is the Jacobian of the inverse transfinite mapping.

To derive the DG scheme, we multiply Eq.~\eqref{eq:1dConservationTransformed} by a locally smooth test function $\phi$ from the same space of $q$ , and integrate over an element to get the weak form:
\begin{equation}
\label{eq: weak form1}
    \int_{-1}^{1} \mathcal{J}\frac{\partial q}{\partial t} \phi d\xi + \int_{-1}^{1} \frac{\partial f_e}{\partial \xi}\phi d\xi = \int_{-1}^{1} \frac{\partial f_v}{\partial \xi}\phi d\xi,\; \forall \phi \in \mathbb{P}_p([-1,1]).
\end{equation}
Considering that $\phi = \sum_{j=0}^{p}c_j\phi_j$ and linear independence between $\phi_j,j=0,\dots,p$, the term associated with inviscid fluxes in Eq.~\eqref{eq: weak form1} can be integrated by parts:

\begin{equation}
\label{eq: weak form2}
    \int_{-1}^{1} \mathcal{J}\frac{\partial q}{\partial t} \phi_j d\xi + f_e\phi_j\Big\vert^{1}_{-1} - \int_{-1}^{1} f_e\frac{\partial \phi_j}{\partial \xi} d\xi = \int_{-1}^{1} \frac{\partial f_v}{\partial \xi}\phi d\xi.
\end{equation}
The solutions between elements are coupled with each other by replacing the discontinuous fluxes at inter-element faces by a numerical inviscid flux, $f^{\ast}_e$, calculated from the Riemann solver\cite{toro_riemann_2009}, which governs the numerical characteristics:
\begin{equation}
    \label{eq: weak form3}
    \int_{-1}^{1} \mathcal{J}\frac{\partial q}{\partial t} \phi_j d\xi + f^{\ast}_e\phi_j\Big\vert^{1}_{-1} - \int_{-1}^{1} f_e\frac{\partial \phi_j}{\partial \xi} d\xi = \int_{-1}^{1} \frac{\partial f_v}{\partial \xi}\phi d\xi.
\end{equation}
In this paper, we adopt the local Lax-Friedrichs (LLF) scheme to compute $f^{\ast}_e$. The viscous fluxes require further manipulations to obtain a usable numerical scheme. Here, we use the Bassi Rebay 1 (BR1) scheme \cite{bassi_high-order_1997}.

For the two-dimensional case, the approach is quite similar. Each element $\Omega_i$ is mapped into a reference element $\Omega_{\text{ref}} = [-1,1]^2$ from the physical space $\vec{x} = (x,y)^{\top}$ to the computational space $\vec{\xi} = (\xi,\eta)^{\top}$ through:
\begin{equation}
    \vec{x} = \vec{X}(\xi,\eta) = X\hat{x} + Y\hat{y}.
\end{equation}
The transformed conservative law is solved on square reference elements:
\begin{equation}
\label{eq:2dconservativelawTransformed}
    \mathcal{J}\frac{\partial \vec{q}}{\partial t} + \frac{\partial\vec{\tilde{f}}_e}{\partial \xi} + \frac{\partial \vec{\tilde{g}}_e}{\partial \eta} = \frac{\partial\vec{\tilde{f}}_v}{\partial \xi} + \frac{\partial \vec{\tilde{g}}_v}{\partial \eta}
\end{equation}
where $\vec{q}$ is the vector of conservative variables. $\tilde{f}$ and $\tilde{g}$ are the contravariant fluxes defined as:
\begin{equation}
    \vec{\tilde{f}} = Y_{\eta}\vec{f} - X_{\eta}\vec{g},\; \vec{\tilde{g}} = -Y_{\xi}\vec{f} + X_{\xi}\vec{g}.
\end{equation}
As in the one-dimensional case, we use the two-dimensional integration by parts to the flux terms and get the weak form of the governing equation:
\begin{equation}
\begin{split}
\label{eq:weak_form_ns}
    \int_{\Omega_{\text{ref}}} \mathcal{J}\frac{\partial \vec{q}}{\partial t} \phi_{ij} d\vec{\xi} & + \oint_{\partial \Omega_{\text{ref}}}\left( \vec{\tilde{f}}_en_{\xi} + \vec{\tilde{g}}_e n_{\eta}\right)\phi_{ij}ds - \int_{\Omega_{\text{ref}}} \vec{\tilde{f}}_e\frac{\partial \phi_{ij}}{\partial \xi} +\vec{\tilde{g}}_e\frac{\partial \phi_{ij}}{\partial \eta}d\vec{\xi} \\
    & = \oint_{\partial \Omega_{\text{ref}}}\left( \vec{\tilde{f}}_vn_{\xi} + \vec{\tilde{g}}_vn_{\eta}\right)\phi_{ij}ds - \int_{\Omega_{\text{ref}}} \vec{\tilde{f}}_v\frac{\partial \phi_{ij}}{\partial \xi} +\vec{\tilde{g}}_v\frac{\partial \phi_{ij}}{\partial \eta}d\vec{\xi}.
\end{split}
\end{equation}
where $\vec{n} = (n_{\xi}, n_{\eta})^\top$ is the normal outward vector of $\partial \Omega_{\text{ref}}$. However, we can integrate the third term by parts once again and replace the flux discontinuous fluxes on interfaces by numerical fluxes to get the strong form:
\begin{equation}
\begin{split}
\label{eq:strong_form_ns}
    \int_{\Omega_{\text{ref}}} \mathcal{J}\frac{\partial \vec{q}}{\partial t} \phi_{ij} d\vec{\xi} &+ \oint_{\partial \Omega_{\text{ref}}}\left[\left( \vec{\tilde{f}}_e^{\ast}- \vec{\tilde{f}}_e\right) n_{\xi} + \left(\vec{\tilde{g}}_e^{\ast} - \vec{\tilde{g}}_e\right)n_{\eta}\right]\phi_{ij}ds + \int_{\Omega_{\text{ref}}} \left( \frac{\partial \vec{\tilde{f}}_e}{\partial \xi} + \frac{\partial \vec{\tilde{g}}_e}{\partial \eta}\right)\phi_{ij}d\vec{\xi} \\
    & = \oint_{\partial \Omega_{\text{ref}}}\left( \vec{\tilde{f}}^{\ast}_vn_{\xi} + \vec{\tilde{g}}^{\ast}_vn_{\eta}\right)\phi_{ij}ds - \int_{\Omega_{\text{ref}}} \vec{\tilde{f}}_v\frac{\partial \phi_{ij}}{\partial \xi} +\vec{\tilde{g}}_v\frac{\partial \phi_{ij}}{\partial \eta}d\vec{\xi}.
\end{split}
\end{equation}
With this in our hands, we can adopt the energy/entropy stable split-form DG\cite{gassner_split_2016} for the inviscid term using Pirozzoli's formulation\cite{pirozzoli_numerical_2011}:
\begin{equation}
    \vec{\tilde{f}}^{\#}_{PI}\left( \vec{q}_{ij},\vec{q}_{kj}\right) = \begin{bmatrix}
        \{\!\{ \rho\}\!\}\{\!\{ u\}\!\} \\
       \{\!\{ \rho\}\!\}\{\!\{ u\}\!\}^2 + \{\!\{ p\}\!\} \\
       \{\!\{ \rho\}\!\}\{\!\{ u\}\!\}\{\!\{ v\}\!\} \\
       \{\!\{ \rho\}\!\}\{\!\{ u\}\!\}\{\!\{ h\}\!\}
       
    \end{bmatrix},
        \vec{\tilde{g}}^{\#}_{PI}\left( \vec{q}_{ij},\vec{q}_{ik}\right) = \begin{bmatrix}
        \{\!\{ \rho\}\!\}\{\!\{ v\}\!\} \\
       \{\!\{ \rho\}\!\}\{\!\{ u\}\!\}\{\!\{ v\}\!\} \\
       \{\!\{ \rho\}\!\}\{\!\{ v\}\!\}^2 + \{\!\{ p\}\!\} \\
       \{\!\{ \rho\}\!\}\{\!\{ v\}\!\}\{\!\{ h\}\!\}
    \end{bmatrix},
\end{equation}
where $\{\!\{ \cdot \}\!\}$ is defined as the mean of two points:
\begin{equation}
    \vec{\tilde{f}}^{\#,1}_{PI}\left( \vec{q}_{ij},\vec{q}_{kj}\right) = \{\!\{ \rho\}\!\}\{\!\{ u\}\!\} = \frac{1}{2}\left( \rho_{ij} + \rho_{kj} \right)\frac{1}{2}\left( u_{ij} + u_{kj} \right),
\end{equation}
and the enthalpy $h = (E+p)/\rho$.

\subsection{Spatio-temporal discretisation}
The one dimensional viscous Burgers' equation and the two-dimensional Navier-Stokes equations can be recovered by taking different conservative variables and corresponding fluxes in Eq.~\eqref{eq:1dConservationTransformed} and Eq.~\eqref{eq:2dconservativelawTransformed}.
We use the nodal DGSEM \cite{kopriva2009implementing} where the solution is approximated on Legendre-Gauss(LG)/Legendre-Gauss-Lobatto(LGL) quadrature points in the reference element. The detailed spatio-temporal discretization is given in \ref{sec: AppendixA}. After discretizing using Lagrange basis functions, a system of Ordinary Differential Equations is obtained:
\begin{equation}
\label{eq: semi-discretisation}
    \frac{d\bm{q}}{dt} = \mathcal{R}(\bm{q};t),
\end{equation}
where $\bm{q}\in \mathcal{M}$ denotes the vector of all the nodal values of the numerical solution, $\mathcal{R}(\bm{q};t)$ is the residual of the discretized equation and $\mathcal{M}$ is the manifold where the solution $\bm{q} $ evolves. In this paper, Williamson's low-storage third-order Runge-Kutta (RK3) scheme \cite{williamson_low-storage_1980} is adopted to march Eq.~\eqref{eq: semi-discretisation} in time, which leads to a time-discretized scheme:
\begin{equation}
    \bm{q}^{n+1} = \mathcal{P}(\bm{q}^{n};\Delta t), 
\end{equation}
where $\bm{q}^n := \bm{q}(t_n), t_n =n\Delta t$, $\mathcal{P}: \mathcal{M} \to \mathcal{M}$ is the discretized time-marching operator parameterized by the time step size $\Delta t$.

\subsection{Corrective Forcing from Neural Networks}\label{sec:force_model}
Given a fixed mesh $\mathcal{T}_h$ in physical space $\Omega$, a high-order simulation (with high polynomial order $p_{ho}$) can be written as follows:
\begin{equation}
\label{eq: ho evolution}
    \bm{q}_{ho}^{n^{\prime}+1} = \mathcal{P}_{ho}(\bm{q}^{n^{\prime}}_{lo};\Delta t_{ho}) ,
\end{equation}
where $\bm{q}_{ho}\in \mathcal{M}_{ho}$, $\mathcal{M}_{ho}$ is the high-order manifold where the high-order solution $\bm{q}_{ho}$ evolves, $n^{\prime}$ is the time step index for $\bm{q}_{ho}$, $\mathcal{P}_{ho}:\mathcal{M}_{ho} \rightarrow \mathcal{M}_{ho}$ is the high-order time-marching operator, and $\Delta t_{ho}$ is the time step for high-order simulation. Similarly, a low-order simulation (with low polynomial order $p_{lo}< p_{ho}$) on $\mathcal{T}_h$ is:
\begin{equation}
\label{eq: lo evolution}
    \bm{q}_{lo}^{n+1} = \mathcal{P}_{lo}(\bm{q}^{n}_{lo};\Delta t_{lo}), 
\end{equation}
where $\bm{q}_{lo}\in \mathcal{M}_{lo}$, $\mathcal{P}_{lo}:\mathcal{M}_{lo} \rightarrow \mathcal{M}_{lo}$ is a low-order time-marching operator, and $\Delta t_{lo}$ is the time step for the low-order simulation ($\Delta t_{lo}> \Delta t_{ho}$).

To obtain a high-order accurate solution in a low-order manifold, we can filter the high-order solution $\bm{q}_{ho}$ into $\mathcal{M}_{lo}$ through a filter:
\begin{equation}
    \label{eq: filter}
    \bm{\bar{q}}_{ho} = \mathcal{F}_{ho}^{lo}(\bm{q}_{ho}),
\end{equation}
where $\bm{\bar{q}}_{ho} \in \mathcal{M}_{lo}$ is the high-order filtered solution and $\mathcal{F}_{ho}^{lo}:\mathcal{M}_{ho} \rightarrow \mathcal{M}_{lo}$ is the filter operator from the high-order to the low-order manifold. The details of the filter operator implementation can be found in \ref{sec: AppendixB}. By applying the filter to the high-order simulation Eq.~\eqref{eq: ho evolution} and using the corresponding low-order time step, the evolution of $\bm{\bar{q}}_{ho}$ is governed by:
\begin{equation}
     \bm{\bar{q}}^{n+1}_{ho} = \mathcal{\bar{P}}_{ho}(\bm{\bar{q}}_{ho}^{n};\Delta t_{lo})
\end{equation}
where $\mathcal{\bar{P}}_{ho}(\cdot\;;\;\cdot)$ is our unknown target because it will generate the same filtered high-order solution by just evolving the solution in the low-order manifold $\mathcal{M}_{lo}$. Assuming $\Delta t_{lo} = m\Delta t_{ho}$, it can be expressed as a combination of the high-order time-marching operator $\mathcal{P}_{ho}(\cdot\;;\;\cdot)$ and the filter operator $\mathcal{F}_{ho}^{lo}(\cdot)$:
\begin{equation}
    \mathcal{\bar{P}}_{ho} = \mathcal{F}_{ho}^{lo} \circ \underbrace{\mathcal{P}_{ho} \circ  \cdots \circ \mathcal{P}_{ho}}_{m \: times}\circ (\mathcal{F}_{ho}^{lo})^{-1} =  \mathcal{F}_{ho}^{lo} \mathcal{P}_{ho}^{m}(\mathcal{F}_{ho}^{lo})^{-1}.
\end{equation}
However, $\mathcal{\bar{P}}_{ho}$ contains the super-resolution reconstruction $(\mathcal{F}_{ho}^{lo})^{-1}$ and the time-marching operator on $\mathcal{M}_{ho}$, $\mathcal{P}_{ho}$, which are unknown. We only know the low-order time-marching operator $\mathcal{P}_{lo} \neq \mathcal{\bar{P}}_{ho}$, but the high-order operator can be recovered by adding a correcting forcing term $\mathcal{S}$ to the low-order operator:
\begin{equation}
    \mathcal{\bar{P}}_{ho} = \mathcal{P}_{lo} + \mathcal{S}.
\end{equation} 
Following \cite{de_lara_accelerating_2022}, we approximate $\mathcal{\bar{P}}_{ho}$ by adding a modeled neural network source term to $\mathcal{P}_{lo}$:
\begin{equation}
\label{eq: P_nn}
    \mathcal{P}_{nn} = \mathcal{P}_{lo} + \mathcal{S}_{nn} \approx \mathcal{\bar{P}}_{ho}
\end{equation}
where $\mathcal{S}_{nn}: \mathcal{M}_{lo} \rightarrow \mathcal{M}_{lo}$ is the corrective forcing modeled by the neural network. So, at the same time, we define a new simulation:
\begin{equation}
\label{eq: q_nn}
    \bm{q}_{nn}^{n+1} = \mathcal{P}_{nn}(\bm{q}^{n}_{nn};\Delta t_{lo}) = \mathcal{P}_{lo}(\bm{q}^{n}_{nn};\Delta t_{lo}) + \mathcal{S}_{nn}(\bm{q}^{n}_{nn};\Delta t_{lo}).
\end{equation}
Once the forcing term $\mathcal{S}_{nn}$ is known at each time step, the filtered high-order solution $\bm{\bar{q}}_{ho}$ can evolve without the need to solve the high-order solution.

\subsection{Methods of training Neural Network}
In what follows, the corrective forcing $\mathcal{S}_{nn}$ is defined as:
\begin{equation}
\label{eq: def_S_nn}
    \mathcal{S}_{nn}(\; \cdot\;;\Delta t ) = \mathcal{N}_{\theta}(\; \cdot\;)\Delta t,
\end{equation}
where $\mathcal{N}_{\theta}$ is the parameterized neural network by $\theta$. To train the neural network in a supervised way, a high-order simulation trajectory $\{\bm{q}_{ho}^{n^{\prime}} \}$ must be filtered into $\{\bm{\bar{q}}_{ho}^{n} \}$ as the ground truth for training. However, if the training process interacts with the numerical solver (low-order time-marching operator $\mathcal{P}_{lo}$), the training methods can be classified as \emph{static} or \emph{interactive}.

\subsubsection{Static training}
To learn the forcing term $\mathcal{S}$ from filtered high-order data, the most straightforward way is to generate $n_d$ training pairs consisting of inputs and outputs.
\begin{equation}
    \left( \bm{\bar{q}}_{ho}^{i}, \mathcal{S}(\bm{\bar{q}}_{ho}^{i}) \right),\; i =0,2,\dots n_d-1,
\end{equation}
where $\mathcal{S}(\bm{\bar{q}}_{ho}^{i})$ can be computed in advance by:
\begin{equation}
    \mathcal{S}(\bm{\bar{q}}_{ho}^{i}) = \bm{\bar{q}}_{ho}^{i+1} - \mathcal{P}_{lo}(\bm{\bar{q}}_{ho}^{i}),
\end{equation}
and $\bm{\bar{q}}_{ho}$ can be obtained from $\bm{q}_{ho}$ through a filter Eq.~\eqref{eq: filter}, as shown in Fig.\ref{fig: schematic diagram}:
\begin{equation}
    \bm{\bar{q}}_{ho}^i = \mathcal{F}_{ho}^{lo}(\bm{q}_{ho}^{mi^{\prime}}),
\end{equation}
where $i^{\prime}$ and $i$ are the time index for a high and low-order solution, $m\in \mathbb{Z}^+$ is defined as the ratio between low and high-order time steps ($m = \Delta t_{lo}/ \Delta t_{ho}$) and $n_d$ is the number of training pairs. Using supervised learning, the loss function can be defined as the average $l_2$ norm error per step:
\begin{equation}
\label{eq: loss_static}
    \mathcal{L} = \frac{1}{n_d-1}\sum_{i=0}^{n_d-2}\sqrt{\int_{\Omega}\left( \mathcal{S}_{nn}(\bm{\bar{q}}_{ho}^{i}) - \mathcal{S}(\bm{\bar{q}}_{ho}^{i})\right)^2d\Omega}.
\end{equation}
and the parameters of the neural network, $\theta$, are trained to minimize the loss function.

\subsubsection{Interactive training}
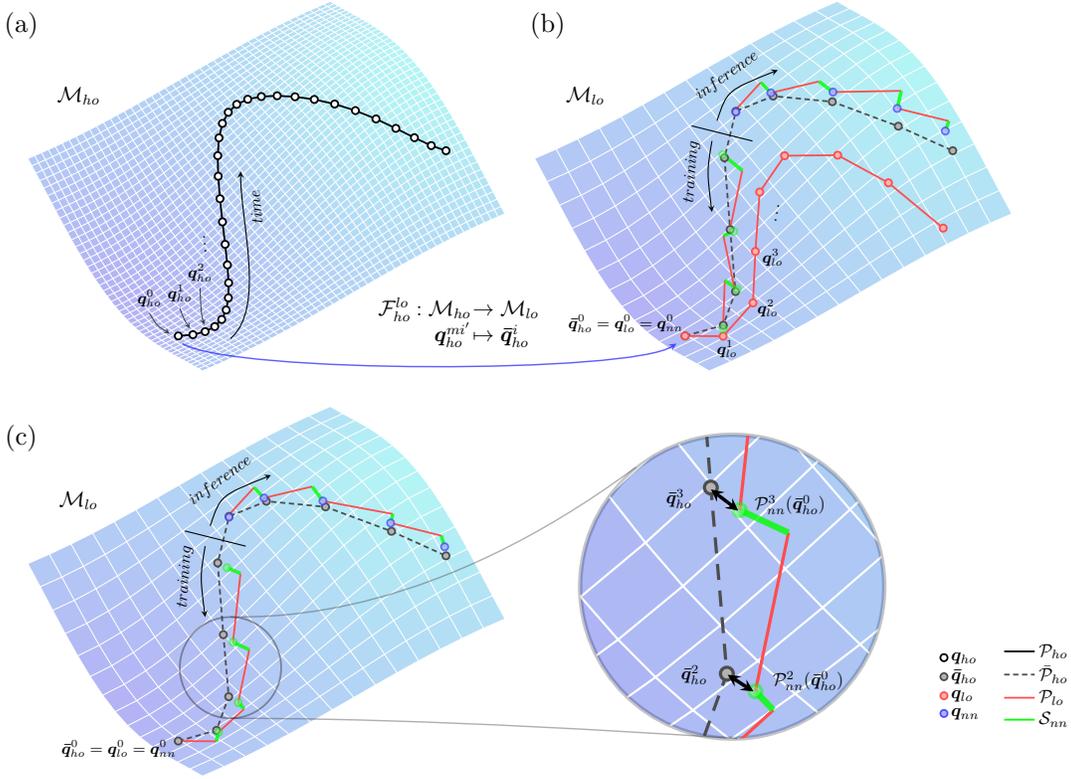
\begin{figure}[h]
	\centering
    \begin{tikzpicture}
        \node[anchor=south west, inner sep=0] (image) at (0,0) {
            \includegraphics[width=0.9\textwidth]{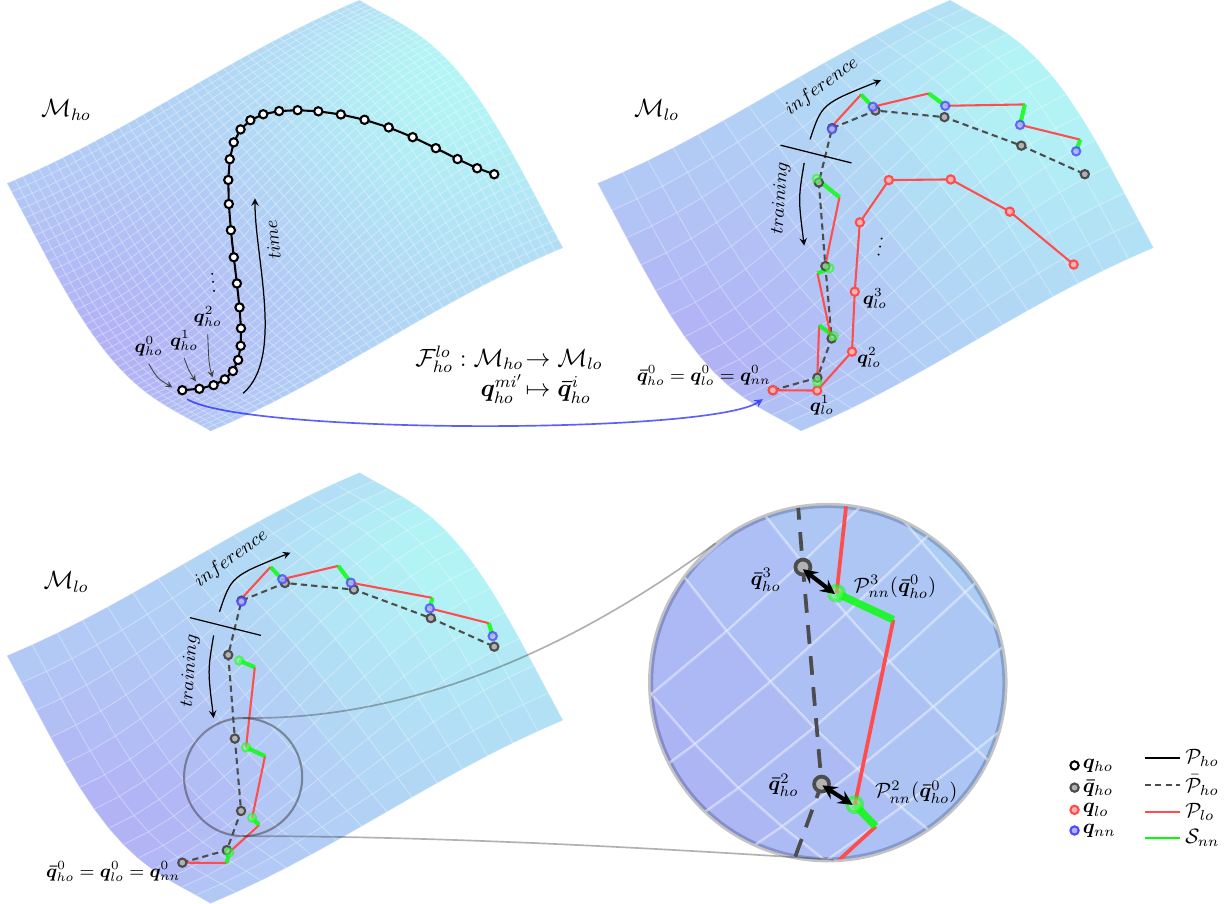}
        };

        \node[black, font=\small, align=center] at (0,10) {(a)};
        \node[black, font=\small, align=center] at (7,10) {(b)};
        \node[black, font=\small, align=center] at (0,4.5) {(c)};
    \end{tikzpicture}
	\caption{Schematic diagram of methodology on accelerating high-order simulation using neural network (a) high-order solution trajectory on $\mathcal{M}_{ho}$; (b) trajectories of low-order simulation, training and inference process on $\mathcal{M}_{lo}$ for $n_{\text{unroll}}=1$ (static training); (c) trajectories of training and inference process on $\mathcal{M}_{lo}$ for $n_{\text{unroll}}=4$ (interactive training).} 
	\label{fig: schematic diagram}
\end{figure}

A problem of static training is the mismatch between the distribution of the training data and the inference data. As shown in Fig.\ref{fig: schematic diagram}(b), all inputs during the training phase are high-order filtered solutions $\bm{\bar{q}}^{i}_{ho}$. However, in the inference phase, the actual input of the neural network is $\bm{q}^{i}_{nn} \neq \bm{\bar{q}}^{i}_{ho}$. As a result, the training error of the neural network will accumulate as time increases because the inputs in the inference phase are different from the distribution in the training phase.

To handle this problem, the inputs of the neural network during the training phase should be as close as possible to that during the inference phase. Generally speaking, the learning target is not to minimize the $l_2$ norm error of a single time integration step in Eq.~\eqref{eq: loss_static}, but to minimize the error of the whole trajectory. Similarly to the definition of loss in \cite{chen_learned_2022}, the loss of a trajectory length of $n_{\text{unroll}}$(starting from the time step $i$) is defined as follows:
\begin{equation}
    \sum_{j=1}^{n_{\text{unroll}}}\sqrt{\int_{\Omega}\left(\mathcal{P}_{nn}^{j}(\bm{\bar{q}}^i_{ho} )- \bm{\bar{q}}_{ho}^{i+j}\right)^{2}d\Omega}.
\end{equation}
where $n_{\text{unroll}}$ is the size of the rolling horizon and $i$ denotes the index of the initial time step of the trajectory. It evaluates the measure of deviation of $\mathcal{P}_{nn}^{j}(\bm{\bar{q}}^i_{ho} )$ from $\bm{\bar{q}}_{ho}^{i+j}$(shown in the zoomed-in part of Fig.~\ref{fig: schematic diagram}(c) and labeled \inlinearrow{}). 

As the total number of high-order filtered solutions $\bm{\bar{q}}_{ho}^{i}$ for training is $n_d$, we can only extract $n_{\text{tr}} = n_d - n_{\text{unroll}}$ trajectories of length of $n_{\text{unroll}}$ from it. To obtain the average $l_2$ norm error per time step, we define the loss function as
\begin{equation}
\label{eq: loss}
    \mathcal{L} = \frac{1}{n_{\text{tr}}n_{\text{unroll}}}\sum_{i=0}^{n_{\text{tr}}-1}\sum_{j=1}^{n_{\text{unroll}}}\sqrt{\int_{\Omega}\left(\mathcal{P}_{nn}^{j}(\bm{\bar{q}}^i_{ho} )- \bm{\bar{q}}_{ho}^{i+j}\right)^{2}d\Omega}.
\end{equation}
Considering that $\mathcal{P}_{nn}(\;\cdot\;)$ is included in the definition, it means that the numerical solver is required to run `online' in the training phase. 

It is worth noting that, in the special case of $n_{\text{unroll}}=1$, the loss function reduces to:
\begin{equation}
\begin{split}
        \mathcal{L} &= \frac{1}{n_d-1}\sum_{i=0}^{n_{d}-2}\sqrt{\int_{\Omega}\left(\bm{q}^{i+1}_{nn}- \bm{\bar{q}}_{ho}^{i+1}\right)^{2}d\Omega} \\
        &= \frac{1}{n_d-1}\sum_{i=0}^{n_{d}-2}\sqrt{\int_{\Omega}\left(\bm{q}^{i+1}_{nn} - \mathcal{P}_{lo}(\bm{\bar{q}}_{ho}^i) - \bm{\bar{q}}_{ho}^{i+1} + \mathcal{P}_{lo}(\bm{\bar{q}}_{ho}^i)\right)^{2}d\Omega} \\
        &= \frac{1}{n_d-1}\sum_{i=0}^{n_d-2}\sqrt{\int_{\Omega}\left( \mathcal{S}_{nn}(\bm{\bar{q}}_{ho}^{i}) - \mathcal{S}(\bm{\bar{q}}_{ho}^{i})\right)^2d\Omega},
\end{split}
\end{equation}
which is exactly the same as Eq.~\eqref{eq: loss_static} in static training because $\bm{q}^i_{nn} = \bm{\bar{q}}^i_{ho}$ when $n_{\text{unroll}} = 1$. The most distinguished feature of $n_{\text{unroll}}>1$ is that the training data input contains not only $\bm{\bar{q}}^{i}_{ho}$, but also the output of $\mathcal{P}_{nn}$, $\bm{q}_{nn}^{i}$. This training strategy enables the neural network to learn its interaction with the solver and integrate with it seamlessly.

To fix ideas, a schematic diagram of the methodology is given in Fig.\ref{fig: schematic diagram}, where static and interactive trainings are sketched. A high-order evolution trajectory $[\bm{q}_{ho}^0,\bm{q}_{ho}^1, \bm{q}_{ho}^2,...]$ is plotted on the high-order manifold $\mathcal{M}_{ho}$ in Fig.\ref{fig: schematic diagram}(a). It is filtered in the low-order manifold $\mathcal{M}_{lo}$ using the filter $\mathcal{F}^{lo}_{ho}:\bm{q}_{ho}^{mi^{\prime}} \mapsto \bm{\bar{q}}_{ho}^{i}$, denoted as gray circles in Figs.\ref{fig: schematic diagram}(b) and (c). 
The training process of $n_{\text{unroll}}=1$ is shown in Fig.\ref{fig: schematic diagram}(b), where $4$ trajectories length of $1$ are used. In contrast, only one trajectory length of $4$ is used to train the neural network for $n_{\text{unroll}}=4$, as shown in Fig.\ref{fig: schematic diagram}(c). Green circles denote the trajectory generated by $\mathcal{P}_{nn}(\;\cdot\;)$ during the training phase. A zoomed-in figure is also shown in Fig.\ref{fig: schematic diagram}(c), where the black double-direction arrow \inlinearrow{} denotes the difference between $\mathcal{P}_{nn}^{j}(\bm{\bar{q}}^i_{ho} )$ and $\bm{\bar{q}}_{ho}^{i+j}$.

\subsubsection{Analysis of propagation of gradients}
Having explained the concepts, we can now give a detailed analysis of the propagation of gradients. Useful gradient information can be computed to update the parameters of the neural network $\theta$ when  a differentiable end-to-end solver is available and $n_{\text{unroll}} >1$, as proposed in this work. Note that we do not need to compute all the gradients below explicitly, because they are evaluated automatically during the training with a differentiable solver. The analysis presented here is to illustrate the propagating process of gradients and to explain the advantage of interactive transmuting over static one.

To compare the difference in gradient propagation between $n_{\text{unroll}}=1$ and $n_{\text{unroll}}>1$ more clearly, here we take $n_{\text{unroll}}=2$ as an example to demonstrate the nature of interactive training. The flow of computation of the first two steps of $\mathcal{P}_{nn}$ is:
\begin{equation*}
    \bm{\bar{q}}_{ho}^0=\bm{q}_{nn}^0 \xrightarrow{\mathcal{P}_{nn}(\;\cdot\;)} \bm{q}_{nn}^1 \xrightarrow{\mathcal{P}_{nn}(\;\cdot\;)} \bm{q}_{nn}^2
\end{equation*}
which is a trajectory length of $2$. Taking into account the definitions of $\mathcal{L}$ in Eq.~\eqref{eq: loss} and $\mathcal{P}_{nn}$ in Eq.~\eqref{eq: P_nn}, its gradients with respect to the parameters of the neural network $\theta$ are:
\begin{equation}
\label{eq: rLrtheta}
    \frac{\partial \mathcal{L}}{\partial \theta} = \frac{\partial \mathcal{L}}{\partial\bm{q}^2_{nn}}\frac{\partial \bm{q}^2_{nn}}{\partial \theta} + \frac{\partial \mathcal{L}}{\partial\bm{q}^1_{nn}}\frac{\partial \bm{q}^1_{nn}}{\partial \theta},
\end{equation}
where $\bm{q}^2_{nn}$ can be written as:
\begin{equation}
\label{eq: qnn2}
    \bm{q}^2_{nn} = \mathcal{P}_{nn}(\bm{q}^1_{nn}) = \mathcal{P}_{lo} (\bm{q}^1_{nn}) + \mathcal{S}_{nn} (\bm{q}^1_{nn}).
\end{equation}
Substituting Eq.~\eqref{eq: qnn2} into Eq.~\eqref{eq: rLrtheta}, we have the following:
\begin{equation}
\begin{split}
    \frac{\partial \mathcal{L}}{\partial \theta} 
    &= \frac{\partial \mathcal{L}}{\partial\bm{q}^2_{nn}}\frac{\partial\left(\mathcal{P}_{lo} (\bm{q}^1_{nn}) + \mathcal{S}_{nn} (\bm{q}^1_{nn})\right)}{\partial \theta} + \frac{\partial \mathcal{L}}{\partial\bm{q}^1_{nn}}\frac{\partial\bm{q}^1_{nn}}{\partial \theta}\\
    \label{eq: rLrtheta2}
    & = \frac{\partial \mathcal{L}}{\partial\bm{q}^2_{nn}}\left(\frac{\partial\mathcal{P}_{lo} (\bm{q}^1_{nn}) }{\partial \bm{q}}\frac{\partial\bm{q}^1_{nn}}{\partial \theta} + \frac{\partial \mathcal{S}_{nn}}{\partial \bm{q}^{1}_{nn}}\frac{\partial \bm{q}^1_{nn}}{\partial \theta} + \frac{\partial\mathcal{S}_{nn}(\bm{q}^1_{nn})}{\partial \theta}\right) + \frac{\partial \mathcal{L}}{\partial\bm{q}^1_{nn}}\frac{\partial\bm{q}^1_{nn}}{\partial \theta}.
\end{split}
\end{equation}
From Eq.~\eqref{eq: q_nn}:
\begin{equation*}
    \frac{\partial\bm{q}^1_{nn}}{\partial \theta} = \frac{\partial \mathcal{P}_{nn}\left( \bm{q}^0_{nn}\right)}{\partial \theta}  = \frac{\partial \mathcal{P}_{lo}( \bm{q}^0_{nn})} {\partial \theta} + \frac{\partial \mathcal{S}_{nn}( \bm{q}^0_{nn})} {\partial \theta},
\end{equation*}
and considering that the low-order solver is independent of $\theta$:
\begin{equation*}
    \frac{\partial \mathcal{P}_{lo}( \bm{q}^0_{nn})} {\partial \theta} = 0,
\end{equation*}
combining the definition of $\mathcal{S}_{nn}$ Eq.~\eqref{eq: def_S_nn}, Eq.~\eqref{eq: rLrtheta2} can be further simplified as:
\begin{equation}
\label{eq: rLrtheta3}
    \frac{\partial \mathcal{L}}{\partial \theta} = \left(\frac{\partial \mathcal{L}}{\partial\bm{q}^2_{nn}}\frac{\partial\mathcal{P}_{lo}}{\partial \bm{q}}\bigg|_{\bm{q}^1_{nn}} + \Delta t_{lo}\frac{\partial \mathcal{N}_{\theta}}{\partial \bm{q}}\bigg|_{\bm{q}^1_{nn}} + \frac{\partial \mathcal{L}}{\partial\bm{q}^1_{nn}} \right)\Delta t_{lo}\frac{\partial\mathcal{N}_{\theta}}{\partial \theta}\bigg|_{\bm{q}^0_{nn}} + \Delta t_{lo}\frac{\partial \mathcal{L}}{\partial\bm{q}^2_{nn}}\frac{\partial\mathcal{N}_{\theta}}{\partial \theta}\bigg|_{\bm{q}^1_{nn}}.
\end{equation}
This expression includes several computations of gradients, and we can classify them into three categories:
\begin{enumerate}
    \item \emph{Loss related}:
    \begin{equation*}
        \frac{\partial \mathcal{L}}{\partial\bm{q}^2_{nn}},\frac{\partial \mathcal{L}}{\partial\bm{q}^1_{nn}},
    \end{equation*}
    they are easy to compute according to the definition of $\mathcal{L}$;

    \item \emph{Neural network related}:
    \begin{equation*}
        \frac{\partial\mathcal{N}_{\theta}}{\partial \theta},\frac{\partial \mathcal{N}_{\theta}}{\partial \bm{q}},
    \end{equation*}
    both can be computed by auto-differentiation because the neural network $\mathcal{N}_{\theta}$ is differentiable from end to end. The first is used to update the parameters $\theta$, while the second term evaluates how the output of the neural network will change as the input $\bm{q}$ changes (can be thought of as the Jacobian of $\mathcal{N}_{\theta}$). Taking ${\partial \mathcal{N}_{\theta}}/{\partial \bm{q}}$ into consideration during training enhances the generalizability of the neural network and the stability of the hybrid simulation;
    
    \item \emph{Solver related}: 
    \begin{equation*}
        J_{lo} = \frac{\partial\mathcal{P}_{lo}}{\partial \bm{q}},
    \end{equation*}
    is the output gradients of the low-order time-marching operator $\mathcal{P}_{lo}$ with respect to its input $\bm{q}$ (the numerical solver gradients). It can be computed using either the AD function in a differentiable solver or the Jacobian matrix of a traditional (non-differentiable) solver. The benefits of adding this term during training is that the interaction between the numerical solver $\mathcal{P}_{lo}$ and the neural network $\mathcal{N}_{\theta}$ can be learned to update the parameters $\theta$ as this gradient is evaluated at $\bm{q} = \bm{q}^1_{nn}$.
\end{enumerate}

If we neglect the terms associated with ${\partial \mathcal{N}_{\theta}}/{\partial \bm{q}}$, ${\partial\mathcal{P}_{lo}}/{\partial \bm{q}}$, and $\bm{q}^1_{nn}$ in Eq.~\eqref{eq: rLrtheta3}, the gradients based on $\mathbf{q}^0_{nn}$ for the simplified method ($n_{\text{unroll}} =1$) reduce to:
\begin{equation}
    \frac{\partial \mathcal{L}}{\partial \theta} = \Delta t_{lo}\frac{\partial \mathcal{L}}{\partial\mathbf{q}^1_{nn}}\frac{\partial\mathcal{N}_{\theta}}{\partial \theta}\bigg|_{\mathbf{q}^0_{nn}}.
\end{equation}
In other words, incorrect gradients are propagated to the neural network in the case of $n_{\text{unroll}}=1$, which only learns the corrective forcing of a single step instead of the entire trajectory.

\subsection{Implementation details}
The energy-stable DGSEM solver is coded using \inlinecode{JAX}\cite{jax2018github}. NN evaluation and training are performed in \inlinecode{JAX} with the use of \inlinecode{Haiku} library \cite{haiku2020github}. All cases are run in a laptop computer with Intel(R) Core(TM) i7-9750H CPU @ 2.60 GHz and RAM 32,0 GB.

\section{Results}\label{sec:results}

In this paper, the benefits of interactive learning through differentiable solver are studied in two cases: the one-dimensional Burger's equation and the two-dimensional Navier-Stokes decaying homogeneous isotropic turbulence (DHIT). Both cases demonstrate that increasing the number of unrolling will enhance the stability of long-term time-stepping, thus increasing the accuracy of the simulation.

\subsection{Case 1: 1D Burger's equation}
\begin{figure}[h]
	\centering
    \hspace*{-1.3cm} 
	\includegraphics[width=0.4\linewidth]{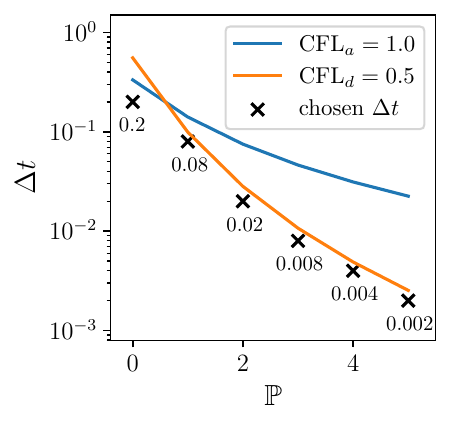} 
	\caption{$\Delta t$ for simulations of different polynomial order $p$. }
	\label{fig: CFL}
\end{figure}

The Burgers' equation is solved in the domain $\Omega = [-1,1]$. Using the same test case in \cite{de_lara_accelerating_2022}, an unsteady boundary condition is imposed on the left end: $q(-1;t) = \sin(1+10t)/2$ and a steady boundary at the right end: $q(1;t) = 1$. The initial condition is $q(x;0) = 1$ at $t = 0$ and the simulation time range is $t \in [0,10]$.
All simulations are performed on the same geometric mesh $\mathcal{T}_h$, where 6 elements are equally distributed within the spatial domain using  a low-storage RK3 temporal scheme. 
For the high-order simulation $\bm{q}_{ho}$, the equation is solved using $\mathbb{P}_5$ elements ($p=5$) while the low-order solution $\bm{q}_{lo}$ is solved on $\mathbb{P}_2$ elements ($p=2$). The time step $\Delta t$ must meet the CFL condition:
\begin{equation}
\label{eq: CFL}
    \Delta t < \min \left\{ \frac{\text{CFL}_a \Delta x_{\min}}{c},\frac{\text{CFL}_d \Delta x^2_{\min}}{\nu} \right\}
\end{equation}
where $\text{CFL}_a$ and $\text{CFL}_d$ are the CFL number for advection and diffusion, $c$ is the wave speed (set to 1) and $\Delta x_{\min}$ is the minimum of space between nodes. The $\Delta t$ computed from Eq.~\eqref{eq: CFL} with $\text{CFL}_a = 1.0$ and $\text{CFL}_d = 0.5$ are plotted in Fig.\ref{fig: CFL}. The $\Delta t$ we choose for different polynomials is labeled in the figure, and the detailed values are also given. For $p_{ho} = 5$ and $p_{lo} = 2$, the ratio between the time steps, $m$, is set equal to $10$.

\begin{figure}[h]
    \centering

    \begin{subfigure}[b]{0.48\textwidth}  
        \centering
        \includegraphics[width=\textwidth]{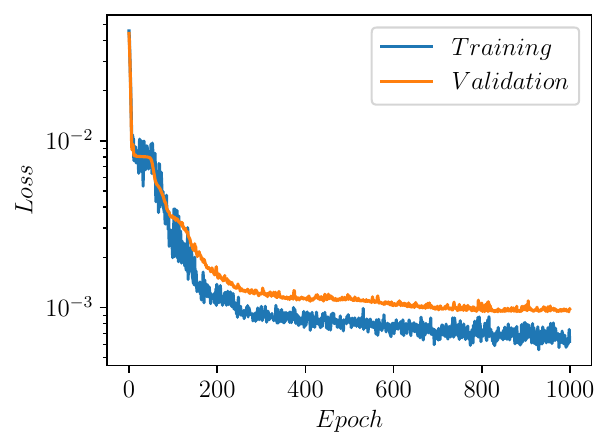}  
        \caption{$n_{\text{unroll}}=1$}
        \label{fig: loss_n_unroll=1}
    \end{subfigure}
    \hfill  
    \begin{subfigure}[b]{0.48\textwidth}
        \centering
        \includegraphics[width=\textwidth]{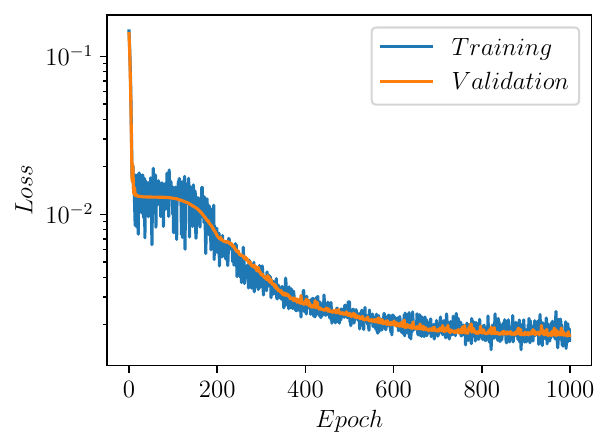}  
        \caption{$n_{\text{unroll}}=5$}
        \label{fig: lossn_unroll=5}
    \end{subfigure}

    \caption{Convergence histories of training and validation loss for two MLPs.}
    \label{fig: loss_MLPs_n_unroll}
\end{figure}

Before training/using NNs, the end-to-end differentiability of the solver is validated in \ref{sec: AppendixC}. Once we are confident in the solver, we compare static and interactive training. For both static and interactive training, the same training strategy is taken to maintain fairness of comparison, except $n_{\text{unroll}} = 1$ for static training and $n_{\text{unroll}} = 5$ for interactive training. Note that various NNs (MLP, CNN, LSTM), which are compared in \cite{Mariño_2024}, showing similar long-term behavior, a very basic neural network type, multilayer perceptron (MLP), is chosen here. The first and last layers are connected to the input and output linearly, 8 hidden layers are connected with $Relu$ activation functions~\cite{sanchez_appendix_2020} and a parabolic distribution of the number of neurons in each layer is used:$[3, 7, 11, 13, 14, 14, 13, 11, 7, 3]$. The entire data are the 50 snapshots of the filtered high-order solution $\bm{\bar{q}}_{ho}$ within the time range $[0,1]$ (out of 10 for the complete simulation), which are obtained by filtering the original high-order solution $\mathbb{P}_5$ $\bm{q}_{ho}$. 70\% of the data are used for training, 20\% are prepared for validation, and the last 10\% are used for testing. The MLPs are trained by 1000 epochs with batch size of 10. The training and validation loss of both MLPs is shown in Fig.~\ref{fig: loss_MLPs_n_unroll}.

\begin{figure}[h]
    \centering
    \begin{tikzpicture}
        \matrix (figgrid) [matrix of nodes, nodes={inner sep=0pt}, row sep=0.2cm, column sep=0.0cm] {
            \includegraphics[height=4.5cm]{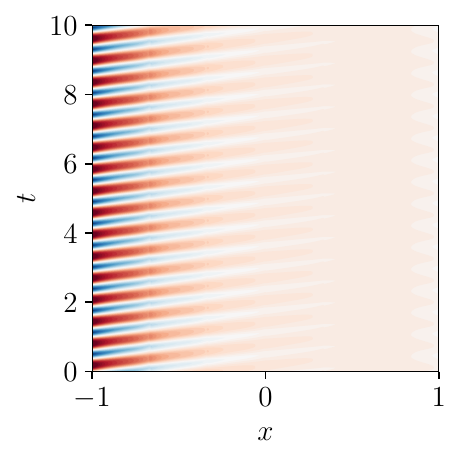} & 
            \includegraphics[height=4.5cm]{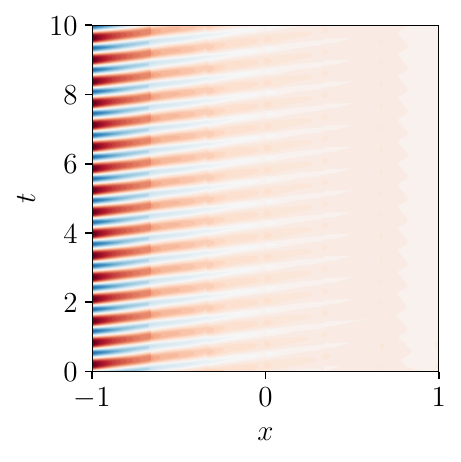} & 
            \includegraphics[height=4.5cm]{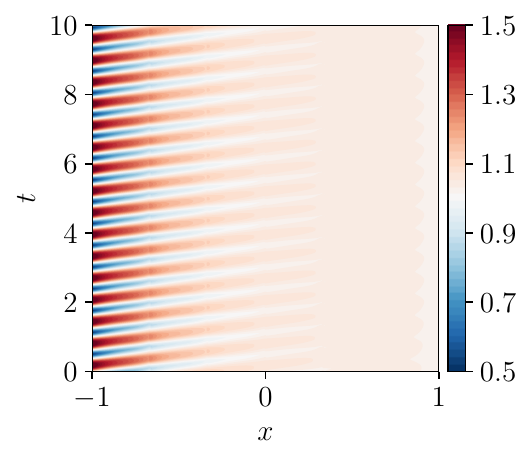} \\ 
        };

        \node[above=-0.2cm, align=center, font=\small] at (figgrid-1-1.north) {$\bm{\bar{q}}_{ho}$};
        \node[above=-0.2cm, align=center, font=\small] at (figgrid-1-2.north) {$\bm{q}_{lo}$};
        \node[above=-0.2cm, align=center, font=\small] at (figgrid-1-3.north) {$\bm{q}_{nn}$};
    \end{tikzpicture}
    \caption{$x-t$ contours of filtered high-order solution $\bm{\bar{q}}_{ho}$, low-order solution $\bm{q}_{lo}$ and low-order solution with NN correction $\bm{q}_{nn}$.}
    \label{fig: comparision of solution contour}
\end{figure}

\begin{figure}[h]
    \centering
    \begin{tikzpicture}
        \matrix (figgrid) [matrix of nodes, nodes={inner sep=0pt}, row sep=0.2cm, column sep=0.0cm] {
            \includegraphics[height=4.1cm]{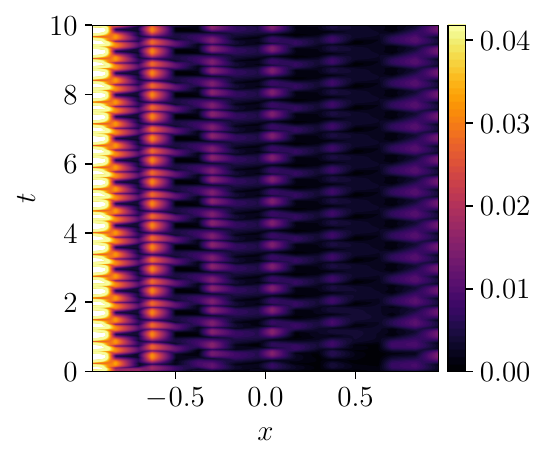} & 
            \includegraphics[height=4.1cm]{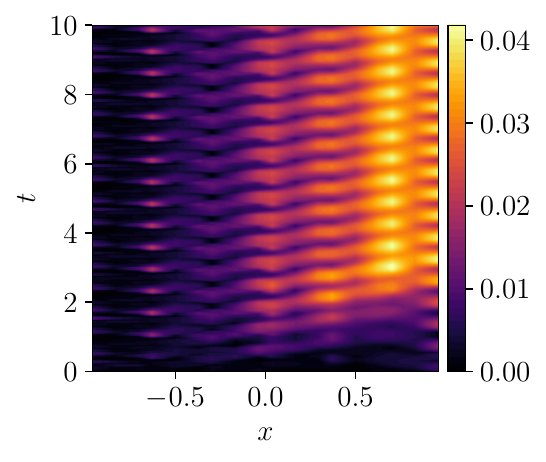} & 
            \includegraphics[height=4.1cm]{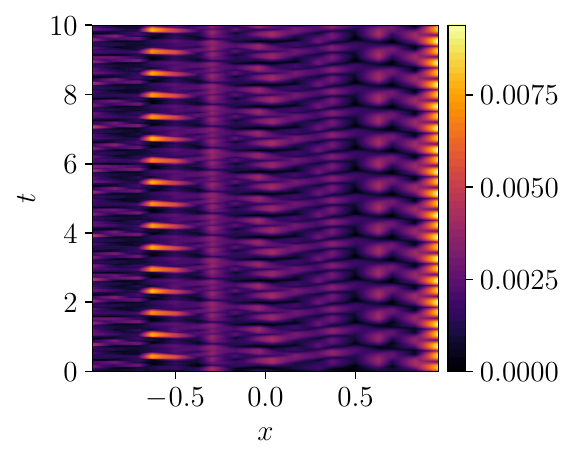} \\ 
        };

        \node[above=-0.2cm, align=center, font=\small] at (figgrid-1-1.north) {$\vert\bm{q}_{lo} - \bm{\bar{q}}_{ho}\vert_1$};
        \node[above=-0.2cm, align=center, font=\small] at (figgrid-1-2.north) {$\vert\bm{q}_{nn} - \bm{\bar{q}}_{ho}\vert_1$($n_{\text{unroll}}=1$)};
        \node[above=-0.2cm, align=center, font=\small] at (figgrid-1-3.north) {$\vert\bm{q}_{nn} - \bm{\bar{q}}_{ho}\vert_1$($n_{\text{unroll}}=5$)};
    \end{tikzpicture}
    \caption{$x-t$ contours of $l_1$ error of low-order solution $\bm{q}_{lo}$, low-order solution with NN correction $\bm{q}_{nn}$($n_{\text{unroll}}=1$ and $n_{\text{unroll}}=5$) compared to filtered high-order solution $\bm{\bar{q}}_{ho}$.}
    \label{fig: error_contour}
\end{figure}

The $x-t$ contour of simulation results of $\bm{\bar{q}}_{ho}$, $\bm{q}_{lo}$ and $\bm{q}_{nn}$ are shown in Fig.~\ref{fig: comparision of solution contour}. It can be clearly seen that there is a weak discontinuity in $\bm{q}_{lo}$ while the solutions of $\bm{\bar{q}}_{ho}$ and $\bm{q}_{nn}$ are much smoother. The difference between $\bm{\bar{q}}_{ho}$ and $\bm{q}_{nn}$ is hard to distinguish, which means that a solution closer to the filtered high-order simulation is recovered by adding the corrective forcing $\mathcal{S}_{nn}$ modeled by the neural network. To illustrate the advantages of interactive training ($n_{\text{unroll}}=5$) over static training ($n_{\text{unroll}}=1$), the space–time evolution of $l_1$ error of $\bm{q}_{lo}$, $\bm{q}_{nn}$ (both $n_{\text{unroll}}=5$ and $n_{\text{unroll}}=1$) compared to $\bm{\bar{q}}_{ho}$ are plotted in Fig.\ref{fig: error_contour}. We observe that the error of $\bm{q}_{nn}$($n_{\text{unroll}}=1$) is a little smaller than that of $\bm{q}_{lo}$, reduced on the left end but increased near the right end. However, $\bm{q}_{nn}$($n_{\text{unroll}}=5$) keeps the error below $1\times10^{-2}$ (the maximum is about $9.2\times 10^{-3}$) in the entire $x-t$ domain, which is much better than in the case $n_{\text{unroll}}=1$.

\begin{figure}[h]
	\centering
    \hspace*{-0.5cm}
	\includegraphics[width=0.9\linewidth]{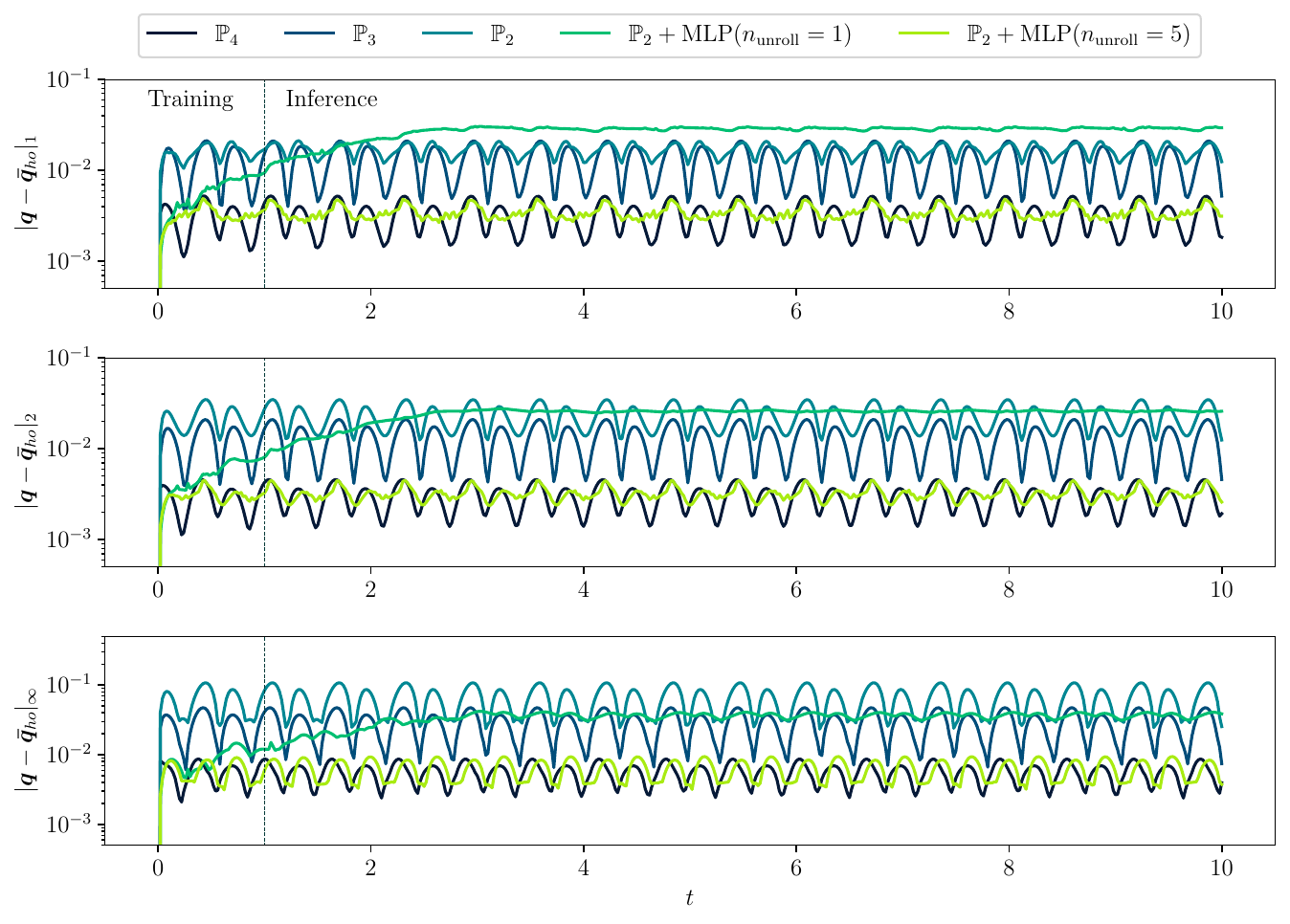} 
	\caption{The evolution of error of different methods.}
	\label{fig: error_evolution}
\end{figure}

To compare the errors $l_1$, $l_2$ and $l_{\infty}$ of different methods, the evolution of the error of the simulations $\mathbb{P}_4$, $\mathbb{P}_3$, $\mathbb{P}_2$ and two $\mathbb{P}_2$ simulations with MLP correction is given in Fig.~\ref{fig: error_evolution}. The $\mathbb{P}_4$ and  $\mathbb{P}_3$ simulations are executed with the time step $ \Delta t$ summarized in Fig.~\ref{fig: CFL}. As the polynomial order $p$ increases from 2 to 4, the error is greatly reduced. With the correction of MLP($n_{\text{unroll}}=1$), the $\mathbb{P}_2$ simulation has a much lower error than without correction in the training phase. However, as the solution evolves, the negative effect of error accumulation can be clearly observed, where the errors are enlarged by almost an order of magnitude. Finally, it cannot be distinguished between the $\mathbb{P}_2$ simulations with and without correction in terms of the error level. In contrast, simulation $\mathbb{P}_2$ with the correction from interactive trained MLP ($n_{\text{unroll}}=5$) maintains a low error level throughout the simulation time range $[0,10]$. The corrective effects in the training and inference phases have the same positive performance, which means that the data shift problem is alleviated by the interactive training through a differentiable solver. All three error norms are comparable to that of $\mathbb{P}_4$. Therefore, we can safely conclude that the accuracy of $\mathbb{P}_4$ is recovered by adding corrective forcing to the $\mathbb{P}_2$ solution. and compute the acceleration ratio without bias. Taking into account $\Delta t_{\mathbb{P}_4} = 4\times10^{-3}$ and $\Delta t_{\mathbb{P}_2} = 2\times10^{-2}$, the $Dofs$ in the time domain is reduced by $\times4$. In the spatial domain, 3 and 5 nodes are used in elements for simulations $\mathbb{P}_2$ and $\mathbb{P}_4$, respectively, which means that the spatial $Dofs$ are reduced by $\times 1.67$. As a result, it can be concluded that $\mathbb{P}_4$ simulation is accelerated by $\times 8.3$ without any loss of accuracy.

\begin{figure}[h]
	\centering
    \hspace*{-0.5cm}
	\includegraphics[width=0.4\linewidth]{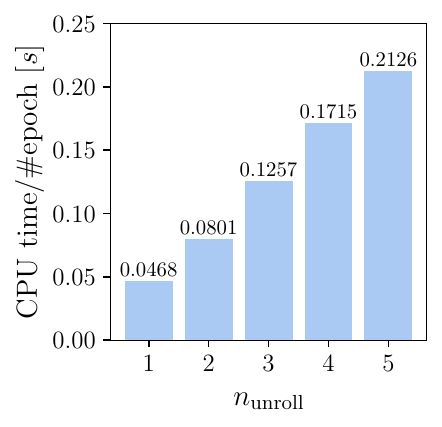} 
	\caption{CPU wall time for training of different $n_{\text{unroll}}$.}
	\label{fig: training_time}
\end{figure}

Despite improved accuracy, the computational cost for training is also increased more for larger $n_{\text{unroll}}$ because a longer computation flow has to be traced and more gradients have to be calculated. The CPU time for training of different $n_{\text{unroll}}$ is plotted in Fig.~\ref{fig: training_time}, where a linear scale law can be observed. The cost of training large $n_{\text{unroll}}$ is quite expensive. However, in \cite{chen_learned_2022} it was reported that initializing the training with $n_{\text{unroll}}=1$ and gradually increasing $n_{\text{unroll}}$ to the target value will reduce training time without losing precision.

\subsection{Case 2: 2D decaying homogeneous isotropic turbulence}

We now explore a Navier-Stokes compressible case. The two-dimensional decaying homogeneous isotropic turbulence (DHIT) is a classical incompressible flow problem in which the kinetic energy decays as time evolves. We simulate the flow in a square $[0,2\pi]^2$ and periodic boundary conditions are applied and follow San and Staples \cite{san_high-order_2012} for the initialization process. The vorticity distribution in the Fourier space, $\hat{\omega}(k)$, is first initialized based on the assumed initial energy spectrum $E_k$:
\begin{equation}
\label{eq:Ek}
    E(k) = \frac{a_s}{2}\frac{1}{k_p}\left( \frac{k}{k_p}\right)^{2s+1}\exp{\left[ -\left( s + \frac{1}{2}\right) \left( \frac{k}{k_p}\right)^2 \right]},
\end{equation}
where $k = \vert \bm{k}\vert = \sqrt{k_x^2+k_y^2}$. The maximum value of the initial energy spectrum occurs at the wavenumber $k_p$ which is assumed to be $k_p = 4$ here. The coefficient $a_s$ normalizes the initial kinetic energy and is given by:
\begin{equation}
    a_s = \frac{(2s + 1)^{s+1}}{2^ss!},
\end{equation}
where $s$ is a shape parameter and we take $s=3$. Using the vorticity-stream function method, the velocity in Fourier space can be recovered:

\begin{align}
    -k^2\hat{\psi}&=-\hat{\omega} \\
    \hat{u}=\imath k_y\hat{\psi},&\; \hat{v}=-\imath k_x\hat{\psi}
\end{align}
where $\hat{\psi}$ is the Fourier transform of the stream function $\psi$: $\hat{\psi}=\mathscr{F}(\psi)$ and $\imath = \sqrt{-1}$. Finally, the physical velocity is obtained using the inverse Fourier transform. The initial conditions of density and pressure are set based on the Mach number:
\begin{equation}
    \rho(x,y,0)=1,\; p(x,y,0)=\frac{1}{\gamma \text{Ma}^2},
\end{equation}
where $\gamma $ is the specific heat ratio and the Mach number is set as $\text{Ma}\approx0.1$ to approximate the incompressibility. The Reynold number is set on the basis of the Taylor microscale $\lambda_{\text{T}}$:
\begin{gather}
    \text{Re}_{\lambda_{\text{T}}}=\frac{\rho u_{\text{rms}}\lambda_{\text{T}}}{\mu},\\
    \lambda_{\text{T}}= \sqrt{\frac{\langle u_{\text{rms}}^2 \rangle}{\langle \left(\partial u /\partial x\right)^2 + \left(\partial u/\partial y\right)^2 +  \left(\partial v /\partial x\right)^2 + \left(\partial v /\partial y\right)^2 \rangle}},
\end{gather}
where $\langle \;\cdot\;\rangle$ denotes the spatial average, $u^2_{\text{rms}}={u^{\prime}}^2+{v^{\prime}}^2$ and $u^{\prime}=u-\langle u \rangle$. More specifically, we set $u_{\text{rms}}\approx1$ and $\lambda_{\text{T}}=0.23387$ and the viscosity coefficient at $\mu=0.003897 $ to ensure that $\text{Re}_{\lambda_{\text{T}}}\approx 60$. 

\begin{figure}[h]
	\centering
	\includegraphics[width=0.5\linewidth]{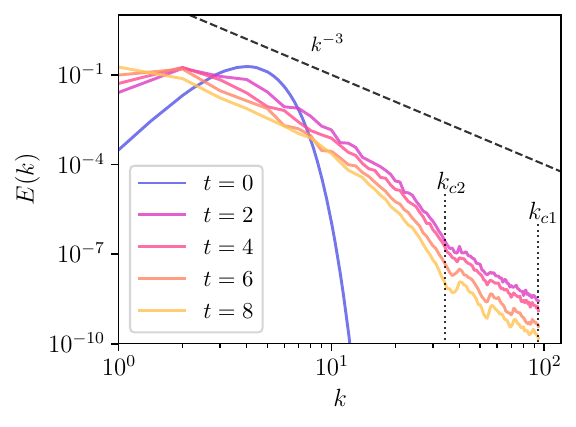} 
	\caption{The energy spectrum of two-dimensional DHIT at $t=0,\;2,\;4,\;6$ and $8$. Two Nyquist wavenumber are computed based on the maximum and the minimum of distance between LGL points ($\Delta x_{\min}$ and $\Delta x_{\max}$):$k_{c_1} = \pi/\Delta x_{min}$, $k_{c_2} = \pi/\Delta x_{max}$. The classical $k^{-3}$ theory is plotted in dash as reference.}
	\label{fig: energy_spectrum}
\end{figure}

We first simulate the flow on $16^2$ $\mathbb{P}_6$ elements using the energy-stable DGSEM \cite{Gassner_2016} with Pirozzoli's two-point flux \cite{pirozzoli_numerical_2011}, LLF Riemann solver and BR1 scheme. The solution is integrated in time using an RK3 scheme. The h-convergence rates of the solver of different polynomial orders are validated using a manufactured analytic solution in \ref{sec: AppendixD}. We check the evolution of the energy spectrum at different times in Fig.~\ref{fig: energy_spectrum}, where the energy spectrum in the inertial range flattens towards the classical $k^{-3}$ scaling, in agreement with the Kraichnan–Batchelor–Leith (KBL) theory of two-dimensional turbulence \cite{kraichnan_inertial_1967,batchelor_computation_1969, leith_atmospheric_1971}. We now simulate the same flow using different polynomial orders, and the corresponding time steps chosen so that the CFL number is approximately 0.25, as listed in Table\ref{table: delta_t_p234567}.

\begin{table}[h]
	\caption{The description of the sub-cases and notation are detailed in the main text.}
    \label{table: delta_t_p234567}
	\begin{center}
		\begin{tabular}{ c | c  c  c  c  c  c} 
                
			\hline
			Polynomial order & $\mathbb{P}_2$ & $\mathbb{P}_3$ & $\mathbb{P}_4$ & $\mathbb{P}_5$ & $\mathbb{P}_6$ & $\mathbb{P}_7$ \\ 
			\hline
		      \rule{0pt}{2.5ex}
              $\Delta t$ & $4\times 10^{-3}$ & $2\times 10^{-3}$ & $1\times 10^{-3}$ & $5\times 10^{-4}$ & $5\times 10^{-4}$ & $4\times 10^{-4}$ \\

            \hline
		\end{tabular}
	\end{center}
\end{table}

\begin{figure}[h]
    \centering

    \begin{subfigure}[b]{0.34\textwidth}  
        \centering
        \adjustbox{valign=t}{\includegraphics[width=\textwidth]{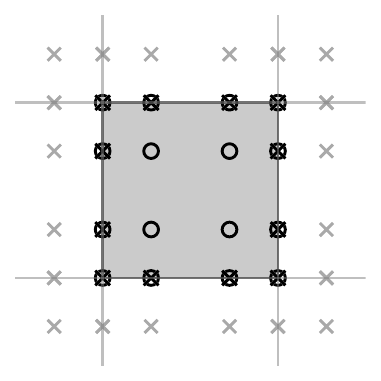}}  
        \caption{}
        \label{fig: inputs_marker}
    \end{subfigure}
    \hfill  
    \begin{subfigure}[b]{0.64\textwidth}
        \centering
        \vspace{0pt} 
        \adjustbox{valign=t}{\includegraphics[width=\textwidth]{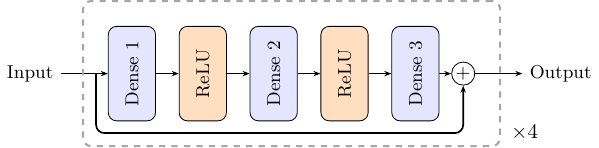}}  
        \vspace{1.4cm} 
        \caption{}
        \label{fig: ResNet_structure}
    \end{subfigure}

    \caption{(a) Schematic diagram of the inputs of NN in a single element($\mathbb{P}_3$ elements are taken as examples here). The element with gray background denotes the current element, where the LGL nodes inside are marked as circles ($\circ$). For the LGL nodes outside current element, they are marked as cross ($\times$). All the black markers denotes the inputs of NN on one element while other gray markers do not. (b) The structure of the NNs (the widths of hidden layers always keep the same as the size of outputs). }
    \label{fig: inputs_and_structure}
\end{figure}

We now study the correction learning approach in three scenarios, which are summarized  in Table~\ref{table: cases summary}. In the first two sub-cases, the baseline low-order simulation is running on $\mathbb{P}_3$ elements and try to learn the correction from the $\mathbb{P}_7$ reference solution. In the third sub-case, we attempt to model the correction from $\mathbb{P}_6$ to $\mathbb{P}_2$. By changing the starting time in Section~\ref{sec: case_2b} to $t=2$, we also study the effectiveness of our approach in different phases of flow-evolution. We trained the NNs to learn the corrections of all four conservative variables. The ResNet \cite{he_deep_2015} framework is adopted in three cases, only changing the sizes of the inputs and outputs. Unlike in our previous work, the inputs of NNs here are not only the nodal values in the current element but also the values on the interfaces of adjacent elements defined via LGL nodes. The schematic figure of the input of NN in a single element is shown in Fig.~\ref{fig: inputs_marker}, where the LGL nodes inside the current element are marked by $\circ$, while the LGL nodes of other adjacent elements are marked by $\times$. All the values on the LGL nodes marked by black are taken as input of NN. It is worth noting that along the boundaries the values from both sides are used because we find that the correction is concentrated on boundaries of elements and highly correlated with the jumps across interfaces. The detailed study on the distribution of correction and its relationship with interfacial jumps can be found in ~\ref{sec: AppendixE}. Therefore, the sizes of the input and output of NN are equal to $4\times (p_{lo}+2)^2$ and $4\times p_{lo}^2$, respectively. The detailed structure of our NN is plotted in Fig.~\ref{fig: ResNet_structure}, where four ResBlocks are used and each block consists of 3 hidden layers. All the training hyper-parameters applied in Case 2 are uniform. NNs are trained up to $1000$ epochs with a batch size of 1 using the Adam optimizer \cite{AdamOptimizer}. The initial learning rate is set equal to 0.01 and decays by 0.97 every 100 epochs. Considering that the NNs contain the modeling of four variables, the total loss function is defined as:
\begin{equation}
    \mathcal{L} = \lambda_1 \mathcal{L}_{\rho} + \lambda_2 \mathcal{L}_{\rho u} + \lambda_3 \mathcal{L}_{\rho v} + \lambda_4 \mathcal{L}_{E},
\end{equation}
where $\mathcal{L}_{\rho},\;\mathcal{L}_{\rho u},\;\mathcal{L}_{\rho v}$ and $\mathcal{L}_{E}$ use the definition of loss in Eq.~\eqref{eq: loss}. $\lambda_i,\;i=1,2,3,4$ are the weight coefficients of the loss of a single variable and here they are set as $\lambda_1 = 10$, $\lambda_2 = 1.0$, $\lambda_3 = 1.0$ and $\lambda_4 = 0.01$ based on the magnitudes of all conservative variables.

\begin{table}[h]
	\caption{Parameters for the studied sub-cases.}
    \label{table: cases summary}
	\begin{center}
		\begin{tabular}{ c | c  c  c } 
                
			\hline
			Sub-cases number & Case 2a & Case 2b & Case 2c \\ 
			\hline
		      Starting time & 0 & 2 & 0\\
            $p_{ho}$ & 7 & 7 & 6 \\ 
            $p_{lo}$ & 3 & 3 & 2 \\
            Training interval & [0,0.08] & [2,2.08] & [0,0.08] \\
            Input size & 144 & 144 & 100 \\ 
            Output size & 64 & 64 & 36 \\
            Tested $n_{\text{unroll}}$ & [1,2,4,8,16]& [1,2,4,8,16]& [1,2,4,8] \\
            \hline
		\end{tabular}
	\end{center}
\end{table}

\subsubsection{Case 2a: $\mathbb{P}_7$ to $\mathbb{P}_3$, starting time $t=0$}
\label{sec: case_2a}

\begin{figure}[h]
	\centering
	\includegraphics[width=1.0\linewidth]{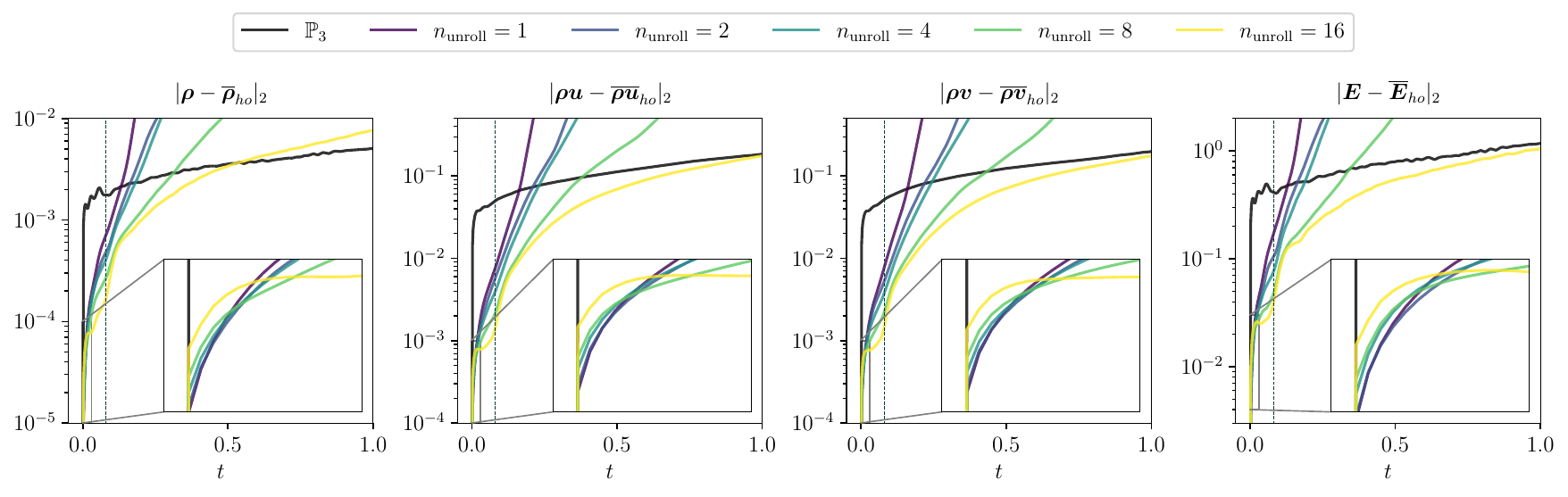} 
	\caption{The evolutions of $l_2$ error of conservative variables of different methods in Case 2a. The vertical dash line denotes the training interval.}
	\label{fig: error_p3_t0-1.0_nn124816}
\end{figure}

\begin{figure}[h]
	\centering
	\includegraphics[width=0.8\linewidth]{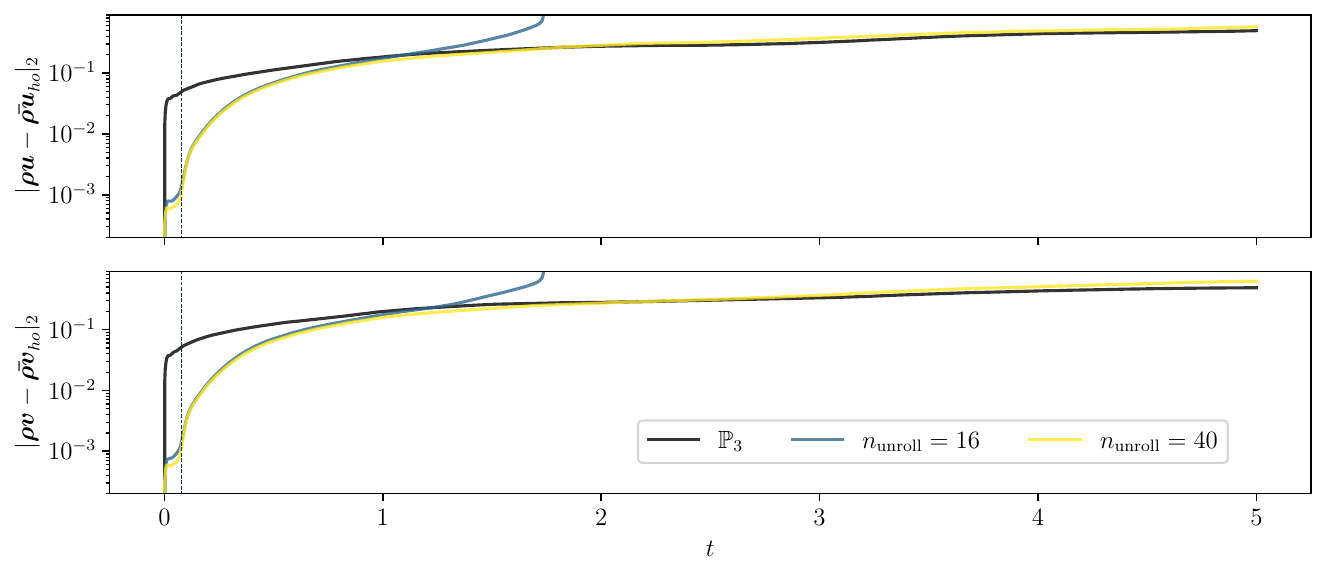} 
	\caption{The evolutions of $l_2$ error of conservative variables of different methods from $t=0$ to $t=5$ in Case 2a.}
	\label{fig: error_p3_t0-5_nn40}
\end{figure}



\begin{figure}[h]
    \centering

    \begin{subfigure}[b]{1.0\textwidth}  
        \centering
        \adjustbox{valign=t}{\includegraphics[width=\textwidth, trim={0cm 0.0cm 0cm 0cm}, clip]{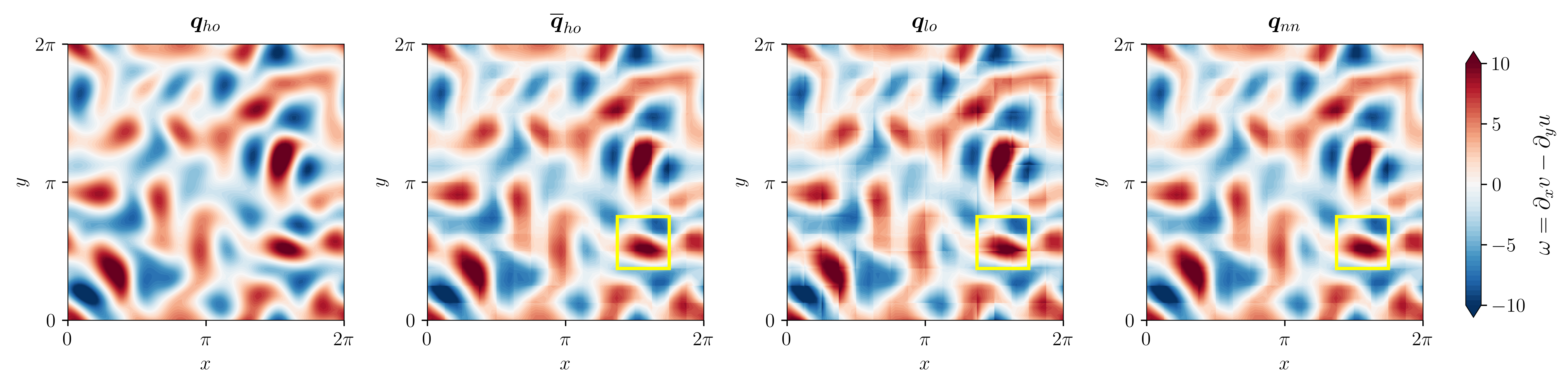}}  
        \caption{Vorticity fields from different methods at $100^{th}$ time step ($t=0.2$). }
        \label{fig: vorticity_case2a_t0.2}
    \end{subfigure}
    \begin{subfigure}[b]{1.0\textwidth}
        \centering
        \adjustbox{valign=t}{\includegraphics[width=\textwidth, trim={0cm 0cm 0cm 0cm}, clip]{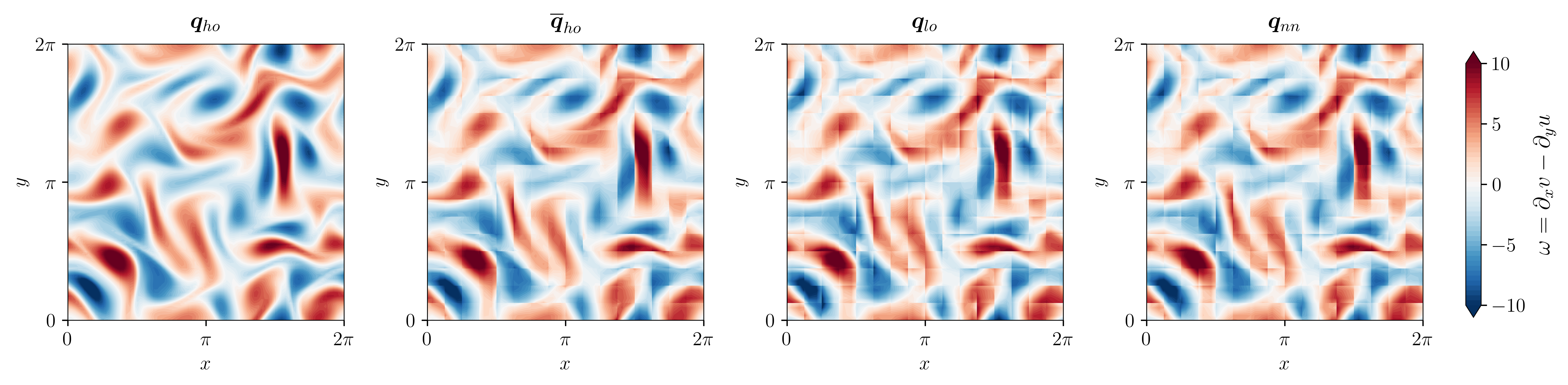}}  
        \caption{Vorticity fields from different methods at $400^{th}$ time step ($t=0.8$).}
        \label{fig: vorticity_case2a_t0.8}
    \end{subfigure}

    \caption{Vorticity fields distribution from $\bm{q}_{ho}$, $\overline{\bm{q}}_{ho}$, $\bm{q}_{lo}$ and $\bm{q}_{nn}$($n_{\text{unroll}}=40$) at (a) $100^{th}$ time step ($t=0.2$) and (b) $400^{th}$ time step ($t=0.8$) for Case 2a.}
    \label{fig: vorticity_case2a}
\end{figure}

\begin{figure}[h]
	\centering
	\includegraphics[width=1.0\linewidth]{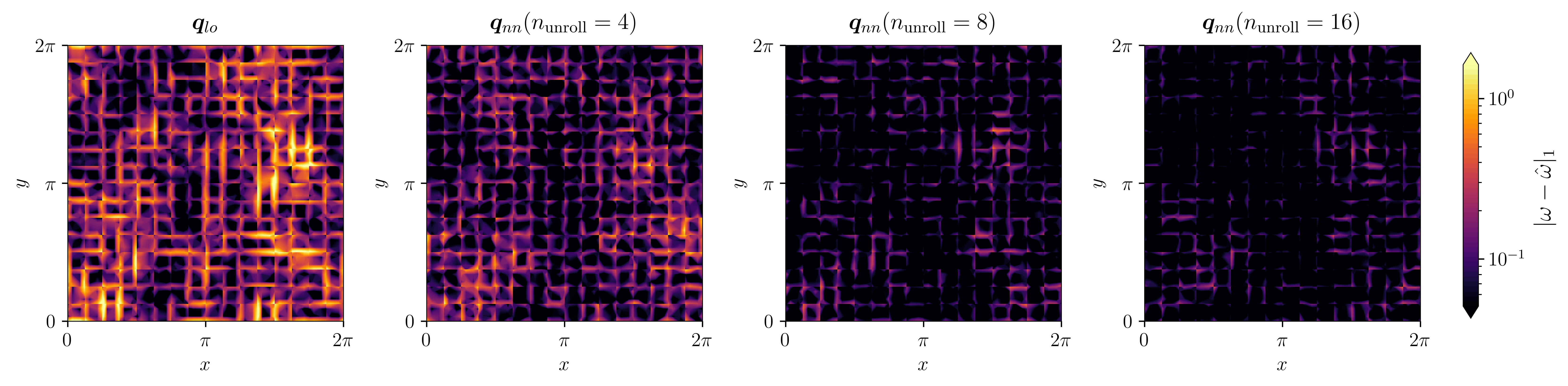} 
	\caption{Vorticity error of $\bm{q}_{lo}$ and $\bm{q}_{nn}$ with $n_{\text{unroll}}=4$, $8$ and $16$ with respect to $\overline{\bm{q}}_{ho}$ (filtered from $\mathbb{P}_7$ solution) at $100^{th}$ step ($t=0.2$).}
	\label{fig: vorticity_error_case2a}
\end{figure}

In this sub-case, we take the filtered high-order simulation from $0$ to $0.08$ ($40$ snapshots) as training data and trained the NNs using different $n_{unroll}$. After that, we run the low-order simulation with learned correction and compute the evolution of $l_2$ error of four conservative variables. As shown in Fig.~\ref{fig: error_p3_t0-1.0_nn124816}, with increasing $n_{\text{unroll}}$, the learned NN model is becoming increasingly stable. For the case of $n_{\text{unroll}}=16$, it reduces the errors of $\rho u$ and $\rho v$ of the original simulation by nearly two magnitudes in the training interval and continues to improve the precision of the low-order simulation up to $t=1$. In contrast, static training ($n_{\text{unroll}}=1$) and other models with small $n_{\text{unroll}}$ are unable to maintain high-accuracy and stability over long periods, whose errors increase very fast. The unrolling size $n_{\text{unroll}}$ can be extended to the length of the entire training data: $n_{\text{unroll}}=40$. The evolution of the error from $t=0$ to $t=5$ is compared in Fig.~\ref{fig: error_p3_t0-5_nn40}. Although the result of $n_{\text{unroll}}=16$ is similar to that of $n_{\text{unroll}}=40$ up to about $t=1$, the simulation with $n_{\text{unroll}}=16$ still suffers from the divergence problem near $t=1.8$. However, correction trained with $n_{\text{unroll}}=40$ enables the simulation to remain stable all the time, reducing the error to $t\approx1.5$ and recovering to the accuracy of the simulation of $\mathbb{P}_3$ when time evolves, which is a desired property of corrective forcing and makes the NN model usable.

The comparison of vorticity from different methods in $t=0.2$ and $t=0.8$ is given in Fig.~\ref{fig: vorticity_case2a}. The original high-order solution $\bm{q}_{ho}$ is also provided as a reference and accurately resolves the vortex. After filtering from $\mathbb{P}_7$ to $\mathbb{P}_3$, some details of the flow are lost, but the discontinuities along the interfaces in the filtered solution $\overline{\bm{q}}_{ho}$ are moderate. However, these discontinuities are greatly amplified after low-order simulations. The target of correction is to solve this problem, and from the vorticity fields of the NN-corrected solution $\bm{q}_{nn}$, it illustrates that a smooth solution close to the reference is obtained by adding corrective forcing. At $t=0.2$, it can be seen from Fig.~\ref{fig: vorticity_case2a_t0.2} that the vorticity of $\bm{q}_{nn}$ is smoother than that of $\bm{q}_{lo}$. For instance, in the yellow zoom-in box, the low-order simulation suffers from a jagged vorticity distribution, while adding the learned correction makes the resolution smoother across the interfaces. Next, we also check the vorticity fields at the $400^{th}$ time step $t=0.8)$ in Fig.~\ref{fig: vorticity_case2a_t0.8} and the result of the NN-corrected simulation ($n_{\text{unroll}}=40$ appears to be closer to the original low-order solution, showing discontinuities along the interfaces, but the resolution of the vortex is still better in some parts. Noting that the flow patterns at $t=0.2$ and $t=0.8$ are quite different, the NN trained with large unrolling size still has the capability to correct the solution to some extent, demonstrating its generalizability to unseen flow states.

To further illustrate the error level of various methods, the vorticity errors in $t=0.2$ are plotted in Fig.~\ref{fig: vorticity_error_case2a}. We show the errors in logaritmic scale to underline the differences. For the low-order solution $\bm{q}_{lo}$, the overall error level is large and concentrated on the interfaces, especially in the regions with large gradients. Increasing $n_{\text{unroll}}$ to $4$, the error is greatly suppressed. When $n_{\text{unroll}}$ increases further to $16$, the error is almost invisible, which means that the learned correction reduces the error of the original simulation by a factor of ten. It proves again that the learned correction enables the simulation to achieve high-order accuracy even beyond several times the training interval.

\begin{figure}[h]
	\centering
	\includegraphics[width=0.5\linewidth]{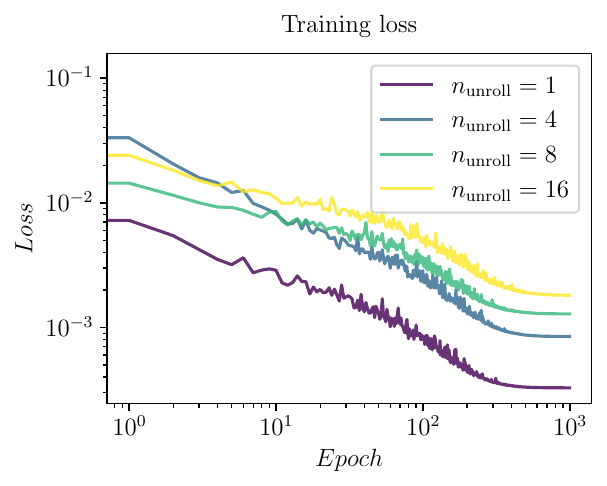} 
	\caption{The comparison of vorticity error distribution of different methods at $100^{th}$ step ($t=0.2$).}
	\label{fig: training_loss_all}
\end{figure}

As was shown for Burgers' equation, a lower training loss in short-time simulation does not ensure better performance over a long period. The histories of training loss for different $n_{\text{unroll}}$ are shown in Fig.~\ref{fig: training_loss_all}. The model with $n_{\text{unroll}}=1$ has the lowest average error per step, as defined in Eq.~\eqref{eq: loss}, and the ultimate loss increases as $n_{\text{unroll}}$ increases. If we check the initial stage (and only for a few steps) in Fig.~\ref{fig: error_p3_t0-1.0_nn124816} (the zoomed-in figure), we see that the error with $n_{\text{unroll}}=1$ is the lowest while the largest error is found for the largest $n_{\text{unroll}}$. However, as time evolves, this trend is reversed:  $n_{\text{unroll}}=1$ is the worst while $n_{\text{unroll}}=16$ provides the best results. This phenomenon can be explained by using the error analysis in \cite{de_lara_accelerating_2022}:
\begin{equation}
    \frac{d\Vert \bm{e}_{nn} \Vert}{dt} \leq \bigg\Vert \frac{\partial \mathcal{P}_{lo}}{\partial \overline{\bm{q}}_{ho}}\bigg\Vert \cdot \Vert \bm{e}_{nn} \Vert + \bigg\Vert \frac{\partial \mathcal{S}_{nn}}{\partial \overline{\bm{q}}_{ho}}\bigg\Vert \cdot \Vert \bm{e}_{nn} \Vert + \Vert \bm{e}_s \Vert,
\end{equation}
where $\bm{e}_{nn} = \overline{\bm{q}}_{ho} - \bm{q}_{nn}$, $\bm{e}_{s} = \mathcal{S} - \mathcal{S}_{nn}$ and $\Vert \;\cdot\; \Vert $ is a general norm. If we approximate the overall effects of the Jacobians of the solver and NN ($\Vert \frac{\partial \mathcal{P}_{lo}}{\partial \overline{\bm{q}}_{ho}}\Vert$ and $\Vert \frac{\partial \mathcal{S}_{nn}}{\partial \overline{\bm{q}}_{ho}}\Vert$) and $\Vert\bm{e}_{s}\Vert$ by $C$ and $\epsilon$, respectively, the expression can be simplified:
\begin{equation}
    \frac{d\Vert \bm{e}_{nn} \Vert}{dt} \approx C\;\Vert \bm{e}_{nn} \Vert + \epsilon,
\end{equation}
and the evolution of error can be approximated as:
\begin{equation}
\label{eq: e_nn}
    \Vert \bm{e}_{nn} \Vert \approx \frac{\epsilon}{C}\left( e^{Ct} - 1 \right).
\end{equation}

The $\epsilon$ parameter can be seen as the initial error in the first step and determines the `starting point' of the entire evolution curve. In addition, $C$ defines the growth rate of the curve and has a negative effect on the initial error. It can be clearly observed from Eq.~\eqref{eq: e_nn} that it is difficult to obtain a corrective model with both low initial error and growth rate simultaneously. In case of NN with $n_{\text{unroll}}=1$, a small training loss ensures the small initial error (low `starting point') but not a small growth rate. In contrast, training with large $n_{\text{unroll}}$ makes a compromise between initial error and growth rate, because the Jacobians of the solver($\Vert \frac{\partial \mathcal{P}_{lo}}{\partial \overline{\bm{q}}_{ho}}\Vert$) and NN ($\Vert \frac{\partial \mathcal{S}_{nn}}{\partial \overline{\bm{q}}_{ho}}\Vert$) are considered during the training process according to the gradient propagation analysis before Eq.~\eqref{eq: rLrtheta3}. Generally speaking, from the results of our numerical tests, a larger $n_{\text{unroll}}$ leads to larger $\epsilon$ and smaller $C$. However, in practice, $\epsilon$ and $C$ could change during the simulation. For example, although the simulation corrected by NN with $n_{\text{unroll}}=16$ remains at a low error level in the training interval, the error increases rapidly beyond this range (near the vertical dashed line), just as $\epsilon$ and $C$ are reset to higher values.

\subsubsection{Case 2b: $\mathbb{P}_7$ to $\mathbb{P}_3$, starting time $t=2$}
\label{sec: case_2b}

\begin{figure}[h]
    \centering
	\includegraphics[width=1.0\linewidth]{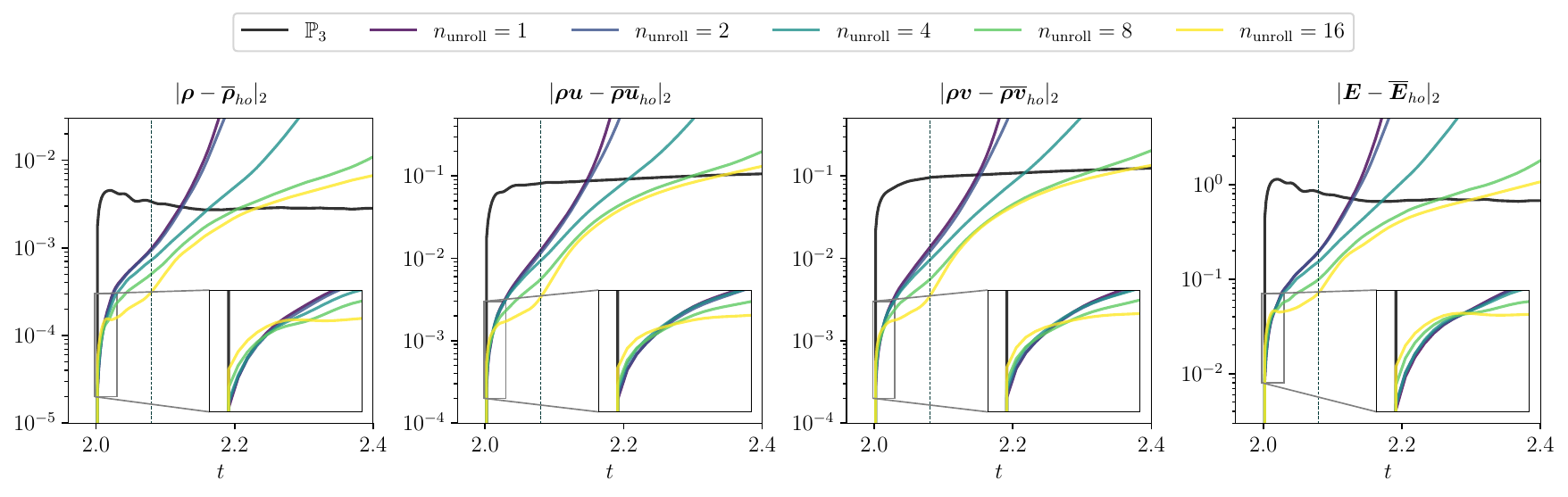} 
	\caption{The evolutions of $l_2$ error of conservative variables of different methods in Case 2b. The time starts at $t=2$.}
	\label{fig: error_p3_t2-5_nn124816}
\end{figure}

\begin{figure}[h]
    \centering

    \begin{subfigure}[b]{1.0\textwidth}  
        \centering
        \adjustbox{valign=t}{\includegraphics[width=\textwidth, trim={0cm 0.3cm 0cm 0cm}, clip]{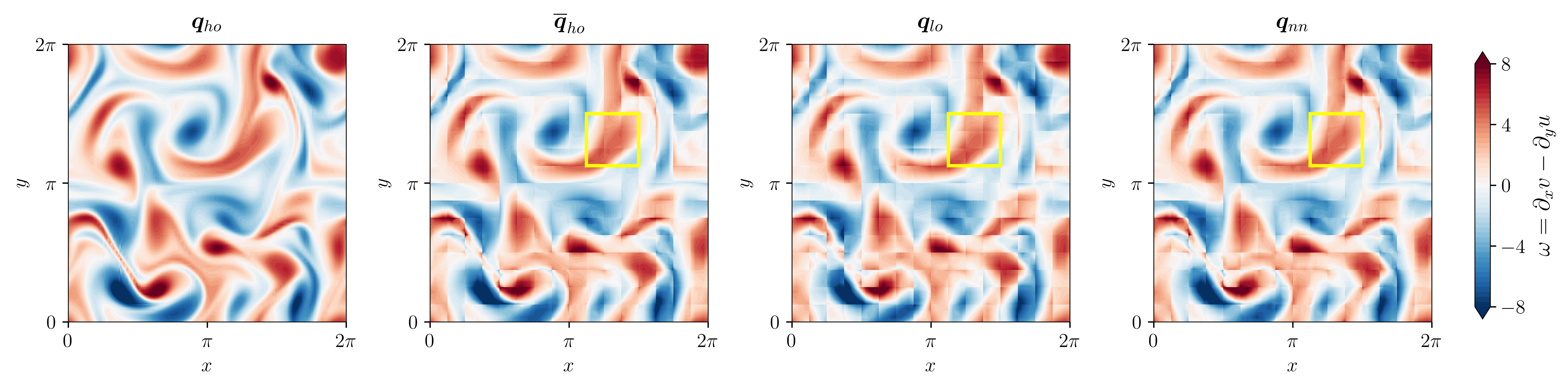}}  
        \caption{Vorticity fields of $\bm{q}_{ho}$, $\overline{\bm{q}}_{ho}$, $\bm{q}_{lo}$ and $\bm{q}_{nn}$($n_{\text{unroll}}=16$). }
        \label{fig: vorticity_case2b}
    \end{subfigure}
    \begin{subfigure}[b]{1.0\textwidth}
        \centering
        \adjustbox{valign=t}{\includegraphics[width=\textwidth, trim={0cm 0.3cm 0.3cm 0cm}, clip]{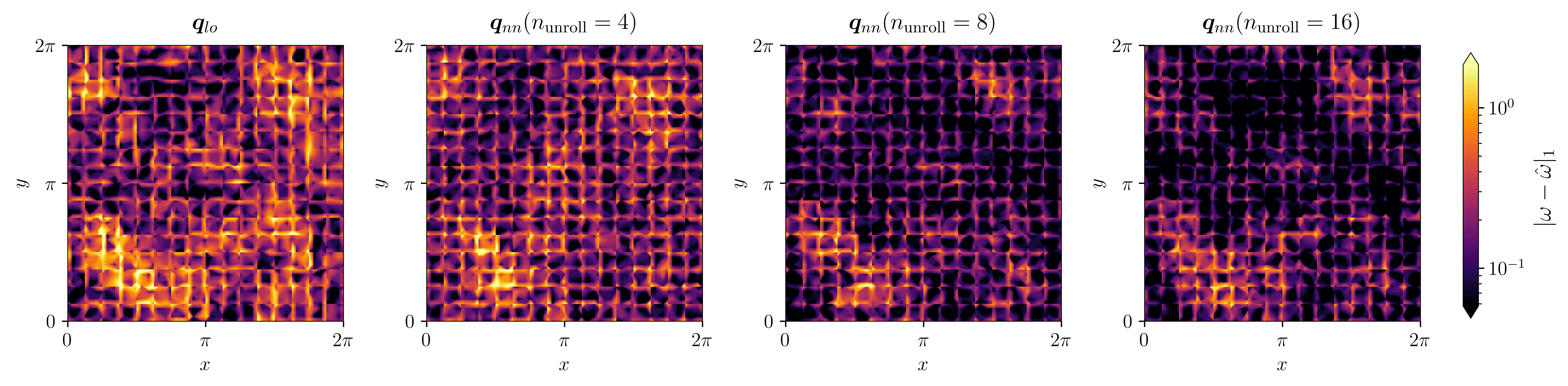}}  
        \caption{Vorticity error of $\bm{q}_{lo}$ and $\bm{q}_{nn}$ with $n_{\text{unroll}}=4$, $8$ and $16$ with respect to $\overline{\bm{q}}_{ho}$ (filtered from $\mathbb{P}_7$ solution).}
        \label{fig: vorticity_error_case2b}
    \end{subfigure}

    \caption{Vorticity fields and error distribution of different methods at $100^{th}$ time step ($t=2.2$) for Case 2b.}
    \label{fig: vorticity_and_error_case2b}
\end{figure}

In this sub-case, we follow the same approach as in Section~\ref{sec: case_2a} except that starting time is shifted to $t=2$ and similar results can be obtained. As shown in Fig.~\ref{fig: error_p3_t2-5_nn124816}, the NN model trained with large $n_{\text{unroll}}$ is much more stable than those trained with small $n_{\text{unroll}}$. However, its effects are not as good as in Case 2a. The potential reason lies in the more complex flow pattern at $t=2$, which enhances under-resolution for low-order simulation. The vorticity distribution of different methods at the $100^{th}$ time step ($t=2.2$) is compared in Fig.~\ref{fig: vorticity_case2b}. It can be seen that the high-order solution $\bm{q}_{ho}$ resolves the flow field well, but the filtered solution $\overline{\bm{q}}_{ho}$ loses some flow details, especially where the vortexes are stretched. The low-order solution $\bm{q}_{lo}$ still has the problem of non-smooth vorticity, and adding the NN-modeled correction mitigates it. As before, the corresponding vorticity errors are shown in Fig.~\ref{fig: vorticity_error_case2b}, and larger $n_{\text{unroll}}$ still provide better results. However, the flow field where vortices are stretched largely suffers from a high error level; for example, in the left-bottom part of the domain. 

\begin{figure}[h]
	\centering
	\includegraphics[width=0.6\linewidth]{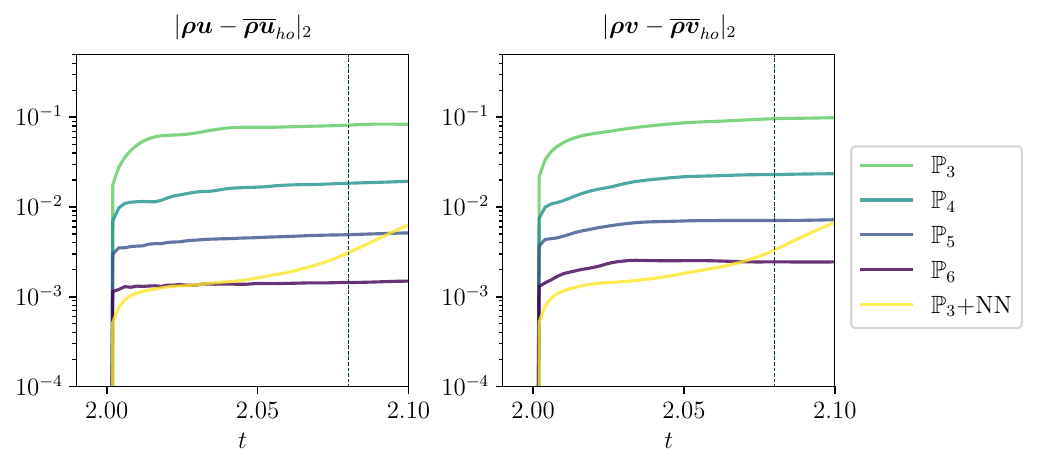} 
	\caption{The comparison of momentum error evolutions of $\bm{q}_{nn}$ and simulations with various polynomial order. Here NN refers to the model trained with $n_{\text{unroll}}=16$.}
	\label{fig: error_p3456_t2-2.1_nn16}
\end{figure}

\begin{figure}[h]
    \centering

    \begin{subfigure}[b]{0.48\textwidth}  
        \centering
        \includegraphics[width=\textwidth]{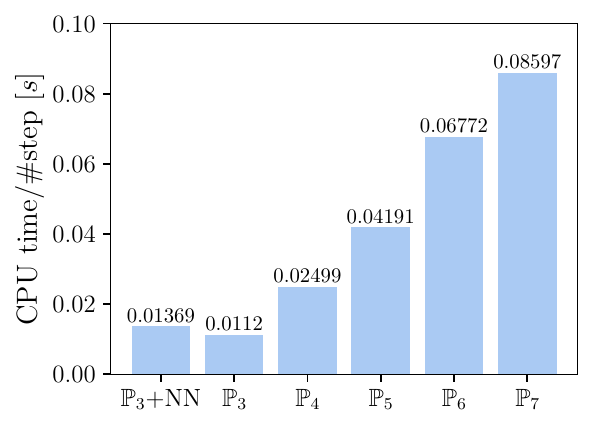}  
        \caption{CPU time per step}
        \label{fig: running_time_per_step}
    \end{subfigure}
    \hfill  
    \begin{subfigure}[b]{0.48\textwidth}
        \centering
        \includegraphics[width=\textwidth]{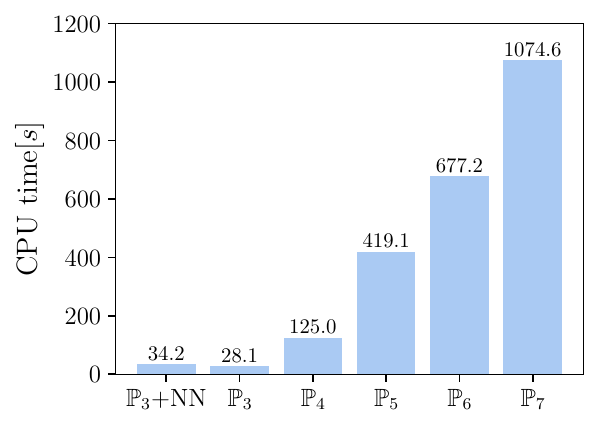}  
        \caption{CPU time of entire simulation}
        \label{fig: running_time_total}
    \end{subfigure}

    \caption{CPU time for NN-corrected and original simulations using different polynomial orders.}
    \label{fig: running_time}
\end{figure}

For completeness, the error evolutions of the NN-corrected method and the simulation with various polynomial orders are compared in Fig.~\ref{fig: error_p3456_t2-2.1_nn16} from $t=2$ to $t=2.1$. In the training interval range, the NN-corrected method performs excellently, maintaining a slow growth rate and achieving high-order simulation precision up to $\mathbb{P}_6$. The correction modeled by NN reduces the error of $\rho u$ and $\rho v$ nearly 100 times. However, the shortcoming lies in that the error grows rapidly in the initial stage beyond the training interval. We also check the computational loads of different methods in Fig.~\ref{fig: running_time}. Figs.~\ref{fig: running_time_per_step} and~\ref{fig: running_time_total} show the CPU time per step and for the entire simulation, respectively. In terms of computational cost per step, high-order simulations are more expensive than low-order ones. However, the additional cost of NN inference occupies only a small percentage of the original method, about $20\%$. The cost of NN-corrected simulation $\mathbb{P}_3$ is much lower than other high-order simulations. This disparity becomes more obvious if the size of the time step is also taken into account, as shown in Fig.~\ref{fig: running_time_total}. 

\subsubsection{Case 2c: $\mathbb{P}_6$ to $\mathbb{P}_2$, starting time $t=0$}
\label{sec: case_2c}

\begin{figure}[h]
    \centering
	\includegraphics[width=1.0\linewidth]{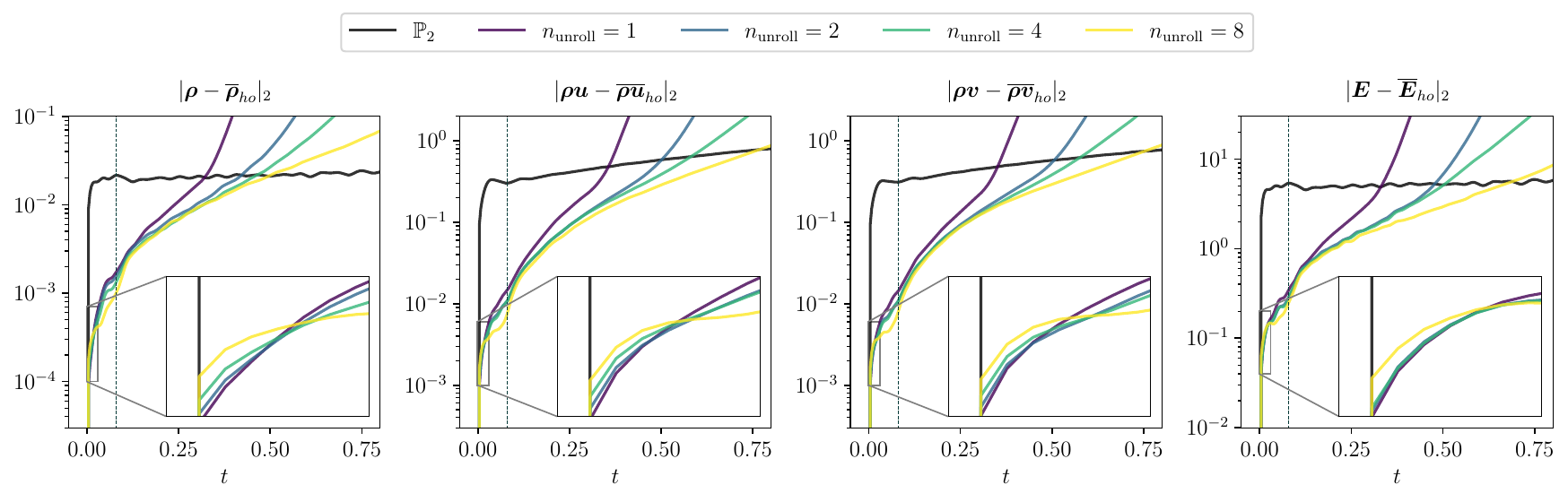} 
	\caption{The evolutions of $l_2$ error of conservative variables of different methods in Case 2c. The time starts at $t=0$.}
	\label{fig: error_p2_t0-0.8_nn1248}
\end{figure}

\begin{figure}[h]
    \centering

    \begin{subfigure}[b]{1.0\textwidth}  
        \centering
        \adjustbox{valign=t}{\includegraphics[width=\textwidth, trim={0cm 0.3cm 0.3cm 0cm}, clip]{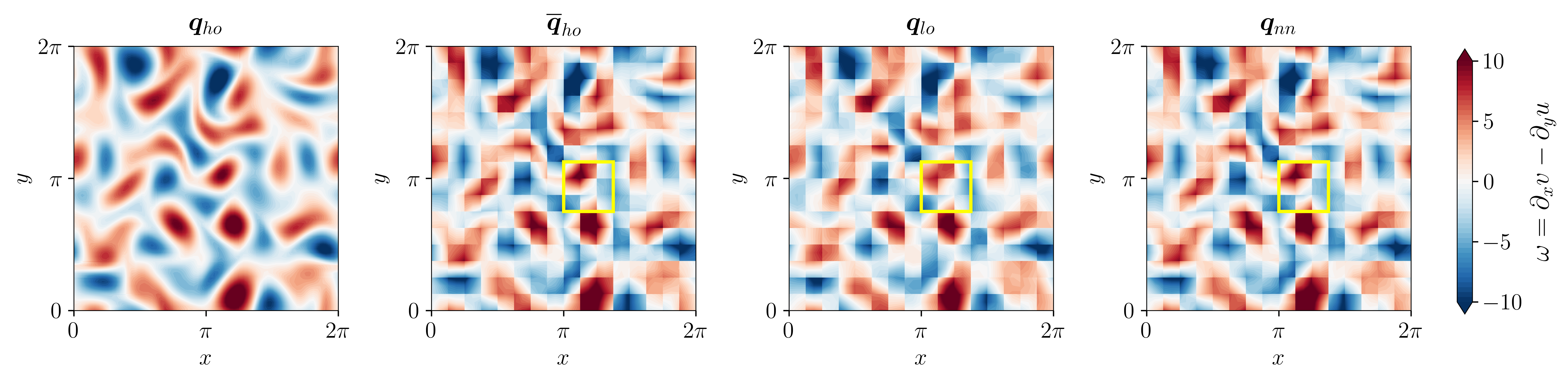}}  
        \caption{Vorticity fields of $\bm{q}_{ho}$, $\overline{\bm{q}}_{ho}$, $\bm{q}_{lo}$ and $\bm{q}_{nn}$($n_{\text{unroll}}=16$). }
        \label{fig: vorticity_case2c}
    \end{subfigure}
    \begin{subfigure}[b]{1.0\textwidth}
        \centering
        \adjustbox{valign=t}{\includegraphics[width=\textwidth, trim={0cm 0.3cm 0cm 0cm}, clip]{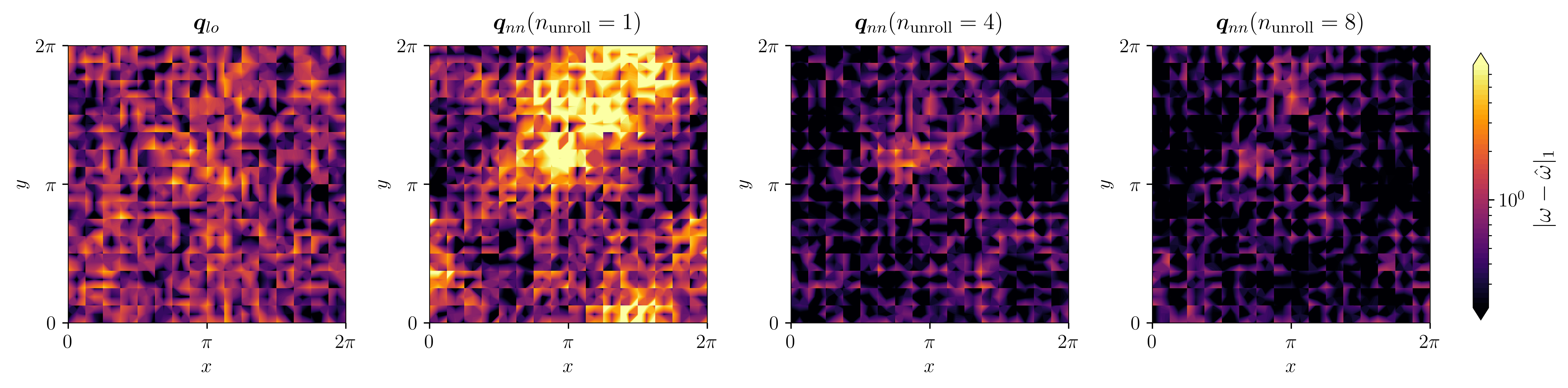}}  
        \caption{Vorticity error of $\bm{q}_{lo}$ and $\bm{q}_{nn}$ with $n_{\text{unroll}}=4$, $8$ and $16$ with respect to $\overline{\bm{q}}_{ho}$ (filtered from $\mathbb{P}_7$ solution).}
        \label{fig: vorticity_error_case2c}
    \end{subfigure}

    \caption{Vorticity fields and error distribution of different methods at $100^{th}$ time step ($t=0.4$) for Case 2c.}
    \label{fig: vorticity_and_error_case2c}
\end{figure}

To further study the effects of the corrective forcing for different polynomial orders, we lower the baseline low-order simulation to $\mathbb{P}_2$ and try to learn the correction from the $\mathbb{P}_6$ solution. Considering that the $\mathbb{P}_2$ mesh is even coarser than the $\mathbb{P}_3$ mesh, the correction forcing should be larger than that on the $\mathbb{P}_3$ elements because the $3^{th}$-order basis is filtered, which contains more energy than other high-order bases. This imposes additional difficulties on the modeling and stability of NN-corrected simulation. However, the $\mathbb{P}_2$ mesh allows for larger $\Delta t$ and provides more numerical dissipation to the simulation, which benefits stability. In addition, smaller input and output sizes ( $64$ and $36$ respectively) lead to fewer trainable parameters in NN, which makes NN training easier. We tested the performance of NNs trained with $n_{\text{unroll}}$ in the range $[1,2,4,8]$ and compared their error evolution in Fig.~\ref{fig: error_p2_t0-0.8_nn1248}. The same as in Case 2a and Case 2b, increasing $n_{\text{unroll}}$ enhances the long-term stability of the NN-corrected simulation. The vorticity fields and the vorticity error of different methods at the $100^{th}$ time step ($t=0.4)$ are plotted in Fig.~\ref{fig: vorticity_and_error_case2c}. It can be observed from Fig.\ref{fig: vorticity_case2c} that $\bm{q}_{ho}$ (from simulation $\mathbb{P}_6$) resolves the flow field well. However, filtering the solution for $\mathbb{P}_2$ leads to a coarser resolution than that of $\mathbb{P}_3$, resulting in loss of detail. Compared to $\overline{\bm{q}}_{ho}$ and $\bm{q}_{nn}$, the vortex structure in $\bm{q}_{lo}$ is highly dissipated; for example, the area in the yellow box. In Fig.~\ref{fig: vorticity_error_case2c}, the result of $n_{\text{unroll}}=1$ is on the edge of divergence, showing a large error due to excessive corrective forcing. As we expected, larger $n_{\text{unroll}}$ leads to a more accurate and stable solution than uncorrected $\bm{q}_{lo}$.

\section{Conclusions}
This work introduces an energy-stable differentiable high-order DGSEM solver that integrates NN-modeled corrective forcing to improve the accuracy of simulation from low-order to high-order precision, thus accelerating the procedure to obtain a high-order solution without compromising accuracy. The end-to-end differentiability allows the back-propagation of gradients through the entire solution trajectory, enabling interactive training strategies (Solver-in-the-Loops) that reduce the data-shift problem and enhance the long-term stability of the simulation. The proposed framework demonstrates a viable path toward solver-informed learning, where physical consistency and numerical stability are embedded in the training process. 

For both Burgers' equation and two-dimensional DHIT cases, the NN-corrected low-order models significantly reduce numerical errors and recover the accuracy of much higher polynomial orders. Interactive training (with unrolling) is shown to outperform static training, yielding smoother gradient propagation, lower error level, and improved generalization to unseen flow states. The influence of unrolling size is studied in the Navier-Stokes two-dimensional decaying homogeneous isotropic turbulence. The numerical tests show that extending the unrolling size generally leads to better long-term stability and accuracy of the simulation. 

Despite the great potential of the method, several challenging problems remain to be overcome to make this method usable. First, the cost of generating high-order solutions as ground-truth data and training NNs is still quite high compared to low-order simulations. Secondly, although the stability of the NN-corrected simulation is greatly enhanced by enlarging the unrolling size, more efforts must be made to improve the inherent properties of the corrective forcing, such as energy-stability, rotation-invariance, and physics consistency. In addition, the design of the NN structure has a large improvement space, and a significant number of advanced ML methods can be applied to complete this learning task.


\section*{Acknowledgments}
Xukun Wang acknowledges the financial support of the China Scholarship Council(CSC, project number: 202406290060). Esteban Ferrer and Oscar Marino acknowledge the funding from the European Union (ERC, Off-coustics, project number 101086075). Views and opinions expressed are, however, those of the authors only and do not necessarily reflect those of the European Union or the European Research Council. Neither the European Union nor the granting authority can be held responsible for them.


\bibliographystyle{elsarticle-num-names} 
\bibliography{refs}

\section*{Appendix}
\appendix

\section{Detailed spatio-temporal discretisation }
\label{sec: AppendixA}

\subsection{1D Burger's equation}
The Burger's equation can be recovered by taking:
\begin{equation}
\label{eq: burgers_eq}
    q = u,\; f_e = \frac{1}{2}u^2,\; f_v = \nu \frac{\partial u}{\partial x}
\end{equation}
in Eq.~\eqref{eq: 1d conservation law}. In DGSEM, the equation to be solved in the reference element $\Omega_{\text{ref}}=[-1,1]$ is given in Eq.~\eqref{eq: weak form3}. Both the solution $q$ and test function $\phi$ are expressed in terms of spectral modal basis $\{\phi_j\}$:
\begin{equation}
\label{eq: modal}
    q(\xi;t) = \sum_{j=0}^{p}\hat{q}_j(t)\phi_j(\xi),
\end{equation}
where $\hat{q}_j$ are the modal coefficients. In this paper, Legendre polynomials $L_j$ are adopted as the spectral basis $\phi_j$. To derive the nodal expression, Eq.~\eqref{eq: modal} can be rewritten as a equivalent Lagrange interpolation form:
\begin{equation}
    q(\xi;t) = \sum_{j=1}^{p+1}q_j\ell_j(\xi),
\end{equation}
where $\ell_j$ are the Lagrange interpolating polynomials and $q_j=q(\xi_j)$ are the nodal values at zeros of $L_j$, $\xi_j$, the Gauss-Legendre quadrature points. Consequently, the Gauss-Legendre quadrature is used to evaluate the integration in weak form, Eq.~\eqref{eq: weak form3}. Therefore, the first term in the left-hand side of Eq.~\eqref{eq: weak form3} can be approximated:
\begin{equation}
\label{eq: int1}
    \int_{-1}^{1}\mathcal{J}\frac{\partial q}{\partial t} \ell_j d\xi \approx \mathcal{J}\sum_{k=1}^{p+1}\left( \sum_{i=1}^{p+1}\frac{dq_i}{dt}\ell_i(\xi_k) \right)\ell_j(\xi_k)w_k.
\end{equation}
Considering the orthogonality property of $\ell_j$: $\ell_i(\xi_j) = \delta_{ij}$, the two-fold summation can be reduced to:
\begin{equation}
\label{eq: Jdqdt}
    \mathcal{J}\frac{dq_j}{dt}w_j. 
\end{equation}
As for the terms associated with $\partial\phi_j/\partial\xi$ in Eq.~\eqref{eq: weak form3}, they can be treated similarly:
\begin{equation}
\label{eq: int2}
    \int_{-1}^{1}f_e\frac{\partial\phi_j}{\partial\xi}d\xi \approx \sum_{k=1}^{p+1}\left( \sum_{i=1}^{p+1}f_{e,i}\ell_i(\xi_k)\right)\ell_j^{\prime}(\xi_k)w_k = \sum_{k=1}^{p+1}f_{e,k}\ell_j^{\prime}(\xi_k)w_k,
\end{equation}
where $\ell_j^{\prime}(\xi_k) = \partial \ell_j/\partial\xi(\xi_k)$ and can be expressed in matrix form:
\begin{equation}
\label{eq: Djk}
    \frac{\partial\ell_j}{\partial\xi}(\xi_k) = D_{kj}.
\end{equation}
Combining Eqs.~\eqref{eq: Jdqdt},~\eqref{eq: int2} and ~\eqref{eq: Djk}, we will obtain the semi-discretisied formulation as Eq.~\eqref{eq: semi-discretisation}:
\begin{equation}
    \mathcal{J}\frac{dq_j}{dt} = -\frac{1}{w_j}\left( f_e^{\ast} - f_d^{\ast}\right)\Big\vert_{-1}^{1}+ \sum_{k=1}^{p+1} \frac{D_{kj}w_k}{w_j}\left( f_{e,k} - f_{d,k}\right), \; j =1,2,\dots,p+1.
\end{equation}
where $f_e^{\ast}$ and $f_e^{\ast}$ are computed through LLF riemann solver and central scheme respectively.

\subsection{2D Navier-Stokes equation}

The original two-dimensional Navier-Stokes equation cane be written in differential form:
\begin{equation}
\label{eq: 2d conservation law}
    \frac{\partial \vec{q}}{\partial t} + \frac{\partial \vec{f}}{\partial x} + \frac{\partial \vec{g}}{\partial y} = \frac{\partial \vec{f}_v}{\partial x} + \frac{\partial \vec{g}_v}{\partial y} + \vec{s}
\end{equation}
where $q$ are conservative variables, $f$ and $g$ are advective flux in $x$- and $y$-directions respectively:
\begin{equation*}
     \vec{q} = \begin{bmatrix}  \rho \\ \rho u \\ \rho v \\ E \end{bmatrix}, \; \vec{f} = \begin{bmatrix}  \rho u \\ \rho u^2 + p \\ \rho uv \\ u(E+p) \end{bmatrix}, \; \vec{g} = \begin{bmatrix}  \rho v \\ \rho vu \\ \rho v^2 + p \\ v(E+p) \end{bmatrix},
\end{equation*}
and the viscous fluxes:
\begin{equation*}
     \vec{f}_v = \begin{bmatrix}  0 \\ \tau_{xx} \\ \tau_{xy} \\ u\tau_{xx} + v\tau_{xy} - q_x \end{bmatrix}, \; \vec{g}_v = \begin{bmatrix}  0 \\ \tau_{yx} \\ \tau_{yy} \\ u\tau_{yx} + v\tau_{yy} - q_y \end{bmatrix}.
\end{equation*}
Assuming a Newtonian fluid and Fourier’s law of thermal conduction yields the stress tensor $\tau$ and heat flux $q$ as:
\begin{equation}
    \begin{split}
        \tau_{xx} &= 2\mu\frac{\partial u}{\partial x} + \lambda\left( \frac{\partial u}{\partial x} + \frac{\partial v}{\partial y}\right),\\
        \tau_{yy} &= 2\mu\frac{\partial v}{\partial y} + \lambda\left( \frac{\partial u}{\partial x} + \frac{\partial v}{\partial y}\right),\\
        \tau_{xy} &= \tau_{yx}=  \mu\left( \frac{\partial u}{\partial y} + \frac{\partial v}{\partial x}\right),\\
        q_x &= -k\frac{\partial T}{\partial x}, q_y = -k\frac{\partial T}{\partial y},
    \end{split}
\end{equation}
where $\mu$ and $\lambda$ are dynamic viscosity and bulk viscosity respectively, and $k$ is the thermal conductivity and $T$ is the temperature. According to the Stokes' hypothesis, $\lambda = -\frac{2}{3}\mu$. Both are material properties of the specific fluid and depend in the general case on the fluid's local state. The thermal conductivity can be computed as:
\begin{equation}
    k = \frac{\gamma R}{\gamma -1} \frac{\mu}{\Pr}
\end{equation}
where $\gamma $ is the ratio of specific heats, $R$ denotes the specific gas constant, and $\Pr$ is the dimensionless Prandtl number, which is assumed in the following to be constant with $Pr = 0.71$.
Finally, the systme is closed the equation of state (EOS):
\begin{equation}
\begin{split}
        p &= (\gamma -1)\left( E - \frac{1}{2}\rho \left(u^2 +v^2\right)\right), \\ 
        & \quad \quad \quad \quad T = \frac{p}{\rho R}.
\end{split}
\end{equation}

Starting from the strong form Eq.~\eqref{eq:strong_form_ns}, we solve the equations on LGL points such that the energy-stable split form DG~\cite{Gassner_2016} can be used. The solution can be expressed by the tensor product of two one-dimensional basis (here we assume that the polynomial order in x- and y- directions are the same):
\begin{equation}
    \vec{q}(\xi,\eta;t)=\sum_{i=1}^{p+1}\sum_{j=1}^{p+1}\vec{q}_{ij}(t)\ell^{(\xi)}_i(\xi)\ell^{(\eta)}_j(\eta),
\end{equation}
and the same for all the fluxes and Jacobian.
\subsubsection{Time derivative term}
By taking $\phi_{ij} = \ell_i^{(\xi)}\ell_j^{(\eta)}$ and considering $\ell_i(\xi_j) = \delta_{ij}$, the time derivative term in Eq.~\eqref{eq:strong_form_ns} becomes:
\begin{equation}
\label{eq:time_derivative}
\begin{split}
    \int_{\Omega_{\text{ref}}} \mathcal{J}\frac{\partial \vec{q}}{\partial t} \phi_{ij} d\vec{\xi} &= \int_{\Omega_{\text{ref}}} \mathcal{J}\frac{\partial \vec{q}}{\partial t} \ell_i^{(\xi)}\ell_j^{(\eta)} d\vec{\xi} \\
    & \approx \sum_{m=1}^{p+1}\sum_{n=1}^{p+1}J_{n,m}\left( \sum_{k=1}^{p+1}\sum_{l=1}^{p+1} \frac{d\vec{q}_{kl}}{dt}\ell_k^{(\xi)}(\xi_m)\ell_l^{(\eta)}(\eta_n)\right)\ell_i^{(\xi)}(\xi_m)^2\ell_j^{(\eta)}(\eta_n)^2\omega_m^{(\xi)}\omega_n^{(\eta)}\\
    & \approx \frac{d\vec{q}_{ij}}{dt} J_{ij}\omega_i^{(\xi)}\omega_j^{(\eta)}
\end{split}
\end{equation}

\subsubsection{Fluxes term}
For the closed surface integration in Eq.~\eqref{eq:strong_form_ns}, it can be written as:
\begin{equation}
\label{eq:surface_term}
    \oint_{\partial \Omega_{\text{ref}}}\vec{\tilde{f}}n_{\xi}\phi_{ij}ds = \int_{-1}^1\phi_{ij}\vec{\tilde{f}} \Big\vert_{\xi = -1}^1d\eta.
\end{equation}
and approximated by:
\begin{align}
    \int_{-1}^1\phi_{ij}\vec{\tilde{f}}_e \Big\vert_{\xi = -1}^1d\eta &\approx \sum_{k = 1}^{p+1} \ell_i^{(\xi)}(1)\ell_j^{(\eta)}(\eta_k)\left(\sum_{m=1}^{p+1}\sum_{n=1}^{p+1}\vec{\tilde{f}}_{e,mn}\ell_m^{(\xi)}(1)\ell_n^{(\eta)}(\eta_k)\right)\omega_k^{(\eta)} \\
    & - \sum_{k = 1}^{p+1} \ell_i^{(\xi)}(-1)\ell_j^{(\eta)}(\eta_k)\left(\sum_{m=1}^{p+1}\sum_{n=1}^{p+1}\vec{\tilde{f}}_{e,mn}\ell_m^{(\xi)}(-1)\ell_n^{(\eta)}(\eta_k)\right)\omega_k^{(\eta)} \\
    & = \ell_i^{(\xi)}(1)\left( \sum_{m=1}^{p+1}\vec{\tilde{f}}_{e,mj}\ell_m^{(\xi)}(1) \right)\omega_j^{(\eta)} - \ell_i^{(\xi)}(-1)\left( \sum_{m=0}^{p+1}\vec{\tilde{f}}_{e,mj}\ell_m^{(\xi)}(-1) \right)\omega^{(\eta)}_j.
\end{align}
Noticing that:
\begin{equation}
    \vec{\tilde{f}}_e(1, \eta_j) = \sum_{m=1}^{p+1}\vec{\tilde{f}}_{e,mj}\ell_m^{(\xi)}(1),\; \vec{\tilde{f}}_e(-1, \eta_j) = \sum_{m=1}^{p+1}\vec{\tilde{f}}_{e,mj}\ell_m^{(\xi)}(-1),
\end{equation}
we will get:
\begin{equation}
\label{eq:term2_1}
    \int_{-1}^1\phi_{ij}\vec{\tilde{f}}_e \Big\vert_{\xi = -1}^1d\eta \approx \left( \ell_i^{(\xi)}(1)\vec{\tilde{f}}_e(1, \eta_j) - \ell_i^{(\xi)}(-1)\vec{\tilde{f}}_e(-1, \eta_j)\right)\omega_j^{(\eta)}.
\end{equation}
Similar operation can be adopted for the surface integration of $\vec{\tilde{g}}_e, \vec{\tilde{f}}^{\ast}_e$ and $\vec{\tilde{g}}^{\ast}_e$ in Eq.~\eqref{eq:strong_form_ns}. Therefore, the surface-related term can be written as:
\begin{equation}
\label{eq:surface-related term}
\begin{split}
    &\oint_{\partial \Omega_{\text{ref}}}\left[\left( \vec{\tilde{f}}_e^{\ast}- \vec{\tilde{f}}_e\right) n_{\xi} + \left(\vec{\tilde{g}}_e^{\ast} - \vec{\tilde{g}}_e\right)n_{\eta}\right]\phi_{ij}ds \\
    \approx &\left(\delta\vec{\tilde{f}}_e(1, \eta_j)\ell_i^{(\xi)}(1) - \delta\vec{\tilde{f}}_e(-1, \eta_j)\ell_i^{(\xi)}(-1)\right)\omega_j^{(\eta)} \\
    +& \left(\delta\vec{\tilde{g}}_e(\xi_i, 1)\ell_j^{(\eta)}(1) - \delta\vec{\tilde{g}}_e(\xi_i, -1)\ell_j^{(\eta)}(-1)\right)\omega_i^{(\xi)}
\end{split}
\end{equation}
where
\begin{equation*}
    \delta\vec{\tilde{f}}_e(\xi,\eta) = \vec{\tilde{f}}_e^{\ast}(\xi,\eta) - \vec{\tilde{f}}_e(\xi,\eta), \; \delta\vec{\tilde{g}}_e(\xi,\eta) = \vec{\tilde{g}}_e^{\ast}(\xi,\eta) - \vec{\tilde{g}}_e(\xi,\eta).
\end{equation*}

Next, the volume-related term in Eq.~\eqref{eq:strong_form_ns} can be approximated similarly:
\begin{equation}
\label{eq:volume-related term}
\begin{split}
    &\int_{\Omega_{\text{ref}}} \left( \frac{\partial \vec{\tilde{f}}_e}{\partial \xi} + \frac{\partial \vec{\tilde{g}}_e}{\partial \eta}\right)\phi_{ij}d\vec{\xi} \\
    &\approx \sum_{m=1}^{p+1}\sum_{n=1}^{p+1}\left( \sum_{k=1}^{p+1}\sum_{l=1}^{p+1}\vec{\tilde{f}}_{e,kl} \frac{\partial  \ell^{(\xi)}_k(\xi_m)}{\partial \xi}\ell^{(\eta)}_l(\eta_n) + \vec{\tilde{g}}_{e,kl} \frac{\partial  \ell^{(\eta)}_l(\eta_n)}{\partial \eta}\ell^{(\xi)}_k(\xi_m)\right)\ell^{(\xi)}_i(\xi_m)\ell^{(\eta)}_j(\eta_n)\omega^{(\xi)}_m\omega^{(\eta)}_n \\
    & = \left( \sum_{k=1}^{p+1}\vec{\tilde{f}}_{e,kj} \frac{\partial  \ell^{(\xi)}_k(\xi_i)}{\partial \xi} + \sum_{l=1}^{p+1}\vec{\tilde{g}}_{e,il} \frac{\partial  \ell^{(\eta)}_l(\eta_j)}{\partial \eta}\right)\omega^{(\xi)}_i\omega^{(\eta)}_j \\
    & = \left( \sum_{k=1}^{p+1} D^{(\xi)}_{ik}\vec{\tilde{f}}_{e,kj} + \sum_{l=1}^{p+1}D^{(\eta)}_{jl}\vec{\tilde{g}}_{e,il} \right)\omega^{(\xi)}_i\omega^{(\eta)}_j
\end{split}   
\end{equation}

To keep entropy stable and energy conservative, I use the split-form DG of \cite{gassner_split_2016} by taking the two-point entropy conserving flux function to approximate the derivatives:
\begin{align}
\label{eq:two-points-entropy}
    \sum_{k=1}^{p+1}D_{ik}^{(\xi)}\vec{\tilde{f}}_{e,kj} &\approx 2 \sum_{k=1}^{p+1}D_{ik}^{(\xi)}\vec{\tilde{f}}^{\#}_{EC}\left( \vec{q}_{ij},\vec{q}_{kj}\right) \\
    \sum_{k=1}^{p+1}D_{jk}^{(\eta)}\vec{\tilde{g}}_{e,ik} &\approx 2 \sum_{k=1}^{p+1}D_{jk}^{(\eta)}\vec{\tilde{g}}^{\#}_{EC}\left( \vec{q}_{ij},\vec{q}_{ik}\right).
\end{align}
where Pirozzoli's formula \cite{pirozzoli_numerical_2011} are chosen:
\begin{equation}
    \vec{\tilde{f}}^{\#}_{PI}\left( \vec{q}_{ij},\vec{q}_{kj}\right) = \begin{bmatrix}
        \{\!\{ \rho\}\!\}\{\!\{ u\}\!\} \\
       \{\!\{ \rho\}\!\}\{\!\{ u\}\!\}^2 + \{\!\{ p\}\!\} \\
       \{\!\{ \rho\}\!\}\{\!\{ u\}\!\}\{\!\{ v\}\!\} \\
       \{\!\{ \rho\}\!\}\{\!\{ u\}\!\}\{\!\{ h\}\!\}
       
    \end{bmatrix},
        \vec{\tilde{g}}^{\#}_{PI}\left( \vec{q}_{ij},\vec{q}_{ik}\right) = \begin{bmatrix}
        \{\!\{ \rho\}\!\}\{\!\{ v\}\!\} \\
       \{\!\{ \rho\}\!\}\{\!\{ u\}\!\}\{\!\{ v\}\!\} \\
       \{\!\{ \rho\}\!\}\{\!\{ v\}\!\}^2 + \{\!\{ p\}\!\} \\
       \{\!\{ \rho\}\!\}\{\!\{ v\}\!\}\{\!\{ h\}\!\}
    \end{bmatrix},
\end{equation}
where $\{\!\{ \cdot \}\!\}$ is defined as the mean of two point:
\begin{equation}
    \vec{\tilde{f}}^{\#,1}_{PI}\left( \vec{q}_{ij},\vec{q}_{kj}\right) = \{\!\{ \rho\}\!\}\{\!\{ u\}\!\} = \frac{1}{2}\left( \rho_{ij} + \rho_{kj} \right)\frac{1}{2}\left( u_{ij} + u_{kj} \right),
\end{equation}
and the enthalpy $h = (E+p)/\rho$.

By treating the viscous term in a similar way and combining Eq.~\eqref{eq:time_derivative}, Eq.~\eqref{eq:surface-related term} and Eq.~\eqref{eq:volume-related term}, we can get the final semi-discretised form:
\begin{equation}
\label{eq:semi-NS}
    \begin{split}
        \frac{d\vec{q}_{ij}}{dt} &+ \frac{1}{J_{ij}}\left\{ \left[\delta\vec{\tilde{f}}_e(1, \eta_j)\frac{\ell_i^{(\xi)}(1)}{\omega_i^{(\xi)}} - \delta\vec{\tilde{f}}_e(-1, \eta_j)\frac{\ell_i^{(\xi)}(-1)}{\omega_i^{(\xi)}}\right] + 2 \sum_{k=1}^{p+1}D_{ik}^{(\xi)}\vec{\tilde{f}}^{\#}_{PI}\left( \vec{q}_{ij},\vec{q}_{kj}\right)\right\} \\
        & + \frac{1}{J_{ij}}\left\{ \left[ \delta\vec{\tilde{g}}_e(\xi_i, 1)\frac{\ell_j^{(\eta)}(1)}{\omega_j^{(\eta)}} - \delta\vec{\tilde{g}}_e(\xi_i, -1)\frac{\ell_j^{(\eta)}(-1)}{\omega_j^{(\eta)}} \right] + 2 \sum_{k=1}^{p+1}D_{jk}^{(\eta)}\vec{\tilde{g}}^{\#}_{PI}\left( \vec{q}_{ij},\vec{q}_{ik}\right) \right\}  \\
        & = \frac{1}{J_{ij}}\left\{ \left[\vec{\tilde{f}}^{\ast}_v(1, \eta_j)\frac{\ell_i^{(\xi)}(1)}{\omega_i^{\xi}} - \vec{\tilde{f}}^{\ast}_v(-1, \eta_j)\frac{\ell_i^{(\xi)}(-1)}{\omega_i^{\xi}}\right] + \sum_{k=1}^{p+1}\hat{D}^{(\xi)}_{ik}\vec{\tilde{f}}_{v,kj}\right\} \\
        & + \frac{1}{J_{ij}}\left\{ \left[ \vec{\tilde{g}}^{\ast}_v(\xi_i, 1)\frac{\ell_j^{(\eta)}(1)}{\omega_j^{\eta}} - \vec{\tilde{g}}^{\ast}_v(\xi_i, -1)\frac{\ell_j^{(\eta)}(-1)}{\omega_j^{\eta}} \right] + \sum_{k=1}^{p+1}\hat{D}^{(\eta)}_{jk}\vec{\tilde{g}}_{v,ik} \right\} \\
        & + \vec{s}_{ij},
    \end{split}
\end{equation}
where the viscous terms are computed using the method by Bassi and Rebay\cite{bassi_high-order_1997}, known as BR1.

\section{Legendre filter implementation}
\label{sec: AppendixB}

The filter operator $\mathcal{F}$ is implemented by applying a uniform Legendre filter to the reference element. The Legendre filter is a truncation of the Legendre series inside the reference element. The nodal values $q_j$ are transferred to Legendre modal coefficients $\hat{q}_{j}$ by projecting solution $q$ onto Legendre polynomials:
\begin{equation}
    \hat{q}_k = \frac{1}{\vert L_k \vert^2}\int_{-1}^1q(\xi)L_k(\xi)d\xi \approx \frac{1}{\vert L_k \vert^2}\sum_{j=1}^{p+1}q_jL_k(\xi_j)w_j.
\end{equation}
Defining the vector of nodal values and modal coefficients of an high-order solution as $\bm{q}_{ho}=[q_1, q_2,\dots,q_{p_{ho}+1}]$ and $\bm{\hat{q}}_{ho}=[\hat{q}_0, \hat{q}_1,\dots,\hat{q}_{p_{ho}}]$ respectively, they satisfy the following equation:
\begin{equation}
\label{eq: transformationmatrix}
    \bm{\hat{q}}_{ho} = \bm{M}_{ho}\bm{q}_{ho}
\end{equation}
where $\bm{M}_{ho}\in \mathbb{R}^{(p_{ho}+1)\times (p_{ho}+1)}$ is the transformation matrix whose elements are defined as:
\begin{equation}
    \bm{M}_{ho}{[i,j]}=\frac{1}{\vert L_{i-1} \vert^2}L_{i-1}(\xi_j)w_j,\; i,j=1,\dots,p_{ho}+1.
\end{equation}
The filtered modal solution $\bm{\bar{\hat{q}}}_{ho}=[\hat{q}_0, \hat{q}_1,\dots,\hat{q}_{p_{lo}-1}]$ is the truncated $\bm{\hat{q}}_{ho}$ by retaining the first $p_{lo}+1$ coefficients, which can be computed directly from $\bm{q}_{ho}$ through:
\begin{equation}
\label{eq: modalfilter}
    \bm{\bar{\hat{q}}}_{ho} = \bm{M}_{ho}^{lo}\bm{q}_{ho}
\end{equation}
where $\bm{M}_{ho}^{lo} \in \mathbb{R}^{(p_{lo}+1)\times (p_{ho}+1)}$ is defined as $\bm{M}_{ho}^{lo} = \bm{M}_{ho}[:p_{lo}+1,:]$. Similar to Eq. ~\eqref{eq: transformationmatrix}, the nodal values of filtered solution on new low-order nodes $\{\xi_j^{\prime}\}$, $\bm{\bar{q}}_{ho}$, and the modal coefficients $\bm{\bar{\hat{q}}}_{ho}$ have the relation:
\begin{equation}
\label{eq: transformationmatrix_lo}
    \bm{\bar{\hat{q}}}_{ho} = \bm{M}_{lo}\bm{\bar{q}}_{ho}
\end{equation}
where $\bm{M}_{lo}\in \mathbb{R}^{(p_{lo}+1)\times (p_{lo}+1)}$ whose definition is the same as $\bm{M}_{ho}$ by replacing $p_{ho}$ by $p_{lo}$. Combining Eq.~\eqref{eq: modalfilter} and~\eqref{eq: transformationmatrix_lo} leads to the final formulation of filter:
\begin{equation}
    \bm{\bar{q}}_{ho} = \bm{M}_{lo}^{-1}\bm{M}_{ho}^{lo}\bm{q}_{ho}
\end{equation}

Considering that two-dimensional basis are the tensor product of two one-dimensional basis, two-dimensional filter can be constructed using the tensor product of two one-dimensional filter:
\begin{equation}
    \bm{\bar{q}}_{ho} = \left[\left(\bm{M}_{lo}^{(\xi)}\right)^{-1}\bm{M}_{ho}^{lo(\xi)}\right] \otimes\left[\left(\bm{M}_{lo}^{(\eta)}\right)^{-1}\bm{M}_{ho}^{lo(\eta)}\right]\bm{q}_{ho}
\end{equation}

\section{Validation of automatic differentiation}
\label{sec: AppendixC}

\begin{figure}[h]
	\centering
	\includegraphics[width=0.6\linewidth]{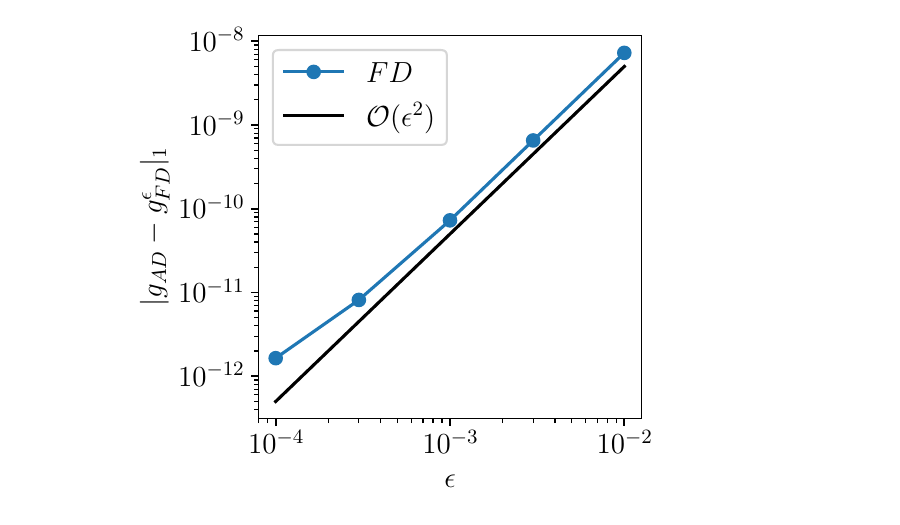} 
	\caption{Error convergence of $g_{FD}^{\epsilon}$ with respective to $g_{AD}$. }
	\label{fig: g_AD-g_FD}
\end{figure}

Before showing the potential of differentiable solver, we first validate the correctness and accuracy of the gradient computed by AD by comparing it with the result of numerical finite difference. To show the end-to-end differentiability of our solver, we solve the burgers' equation Eq.~\eqref{eq: burgers_eq} from a sinusoidal initial condition:
\begin{equation}
    q_0(x) = -sin(\pi x),\; x \in [-1,1]
\end{equation}
with $\nu=0.1$. The domain is discretized by six $\mathbb{P}_5$ elements, and RK3 time integrator is applied with $\Delta t=1.0\times10^{-4}$. The initial condition is integrated by $5$ steps to get the solution $\bm{q}(t_5)$ and its energy is defined as:
\begin{equation}
    E(\bm{q}(t);\nu) = \frac{1}{2}\int_\Omega \bm{q}^2(t;\nu)dx.
\end{equation}
As $\nu$ is a parameter in Burgers' equation, we can compute the gradient of $E(\bm{q}(t_5);\nu)$ with respective to $\nu$ at $\nu =0.1$:
\begin{equation}
    g = \frac{\partial E(\bm{q}(t_5);\nu)}{\partial \nu}\Big\vert_{\nu = 0.1}.
\end{equation}
This can be calculated numerically by a finite difference method (FD) and using central differences:
\begin{equation}
    \label{eq: numerical grandient}
    g_{FD}^{\epsilon} = \frac{E(\bm{q}(t_5);\nu + \epsilon) - E(\bm{q}(t_5);\nu - \epsilon)}{2\epsilon}.
\end{equation}
This formulation has a second-order accuracy, which means Eq.~\eqref{eq: numerical grandient} should converge to the exact value with the speed of $\mathcal{O}(\epsilon^2)$. As shown in Fig.~\ref{fig: g_AD-g_FD}, the second-order convergence of $g_{FD}^{\epsilon}$ to $g_{AD}$ can be observed, where $g_{AD}$ denotes the gradient computed by AD. For the numerical gradient, we compute it by choosing $\epsilon \in [1\times10^{-2},3\times10^{-3},1\times10^{-3},3\times10^{-4},1\times10^{-4}]$. It can be concluded that AD works through the entire solver and computes the gradient with high accuracy.

\section{Convergence rate test}
\label{sec: AppendixD}
We test the h-convergence behavior through a manufactured 2D Euler solution:
\begin{equation}
    \begin{split}
        \rho(x,y,t) &= 2 +\frac{1}{10}\sin\left( \pi (x+y-t)\right),\\
        u(x,y,t) &= 1,\\
        v(x,y,t) &= 1,\\
        E(x,y,t) &= \left(2 +\frac{1}{10}\sin\left( \pi (x+y-t)\right)\right)^2,
    \end{split}
\end{equation}
with the corresponding forcing terms:
\begin{equation}
    \begin{split}
        s_{\rho}(x,y,t) &= c_1 \cos\left( \pi (x+y-t)\right),\\
        s_{\rho u }(x,y,t) &= c_2 \cos\left( \pi (x+y-t)\right) + c_3 \sin\left( 2\pi (x+y-t)\right),\\
        s_{\rho v }(x,y,t) &= c_2 \cos\left( \pi (x+y-t)\right) + c_3 \sin\left( 2\pi (x+y-t)\right),\\
        s_{E }(x,y,t) &= c_4 \cos\left( \pi (x+y-t)\right) + c_5 \sin\left( 2\pi (x+y-t)\right),
    \end{split}
\end{equation}
where $c_1 = \frac{1}{10}\pi$, $c_2 = \frac{1}{10}(3\gamma-2)\pi$, $c_3 = \frac{1}{100}(\gamma-1)\pi$, $c_4 = \frac{1}{5}(3\gamma-2)\pi$ and $c_5 = \frac{1}{100}(2\gamma-1)\pi$. 

\begin{figure}[h]
	\centering
	\includegraphics[width=0.5\linewidth]{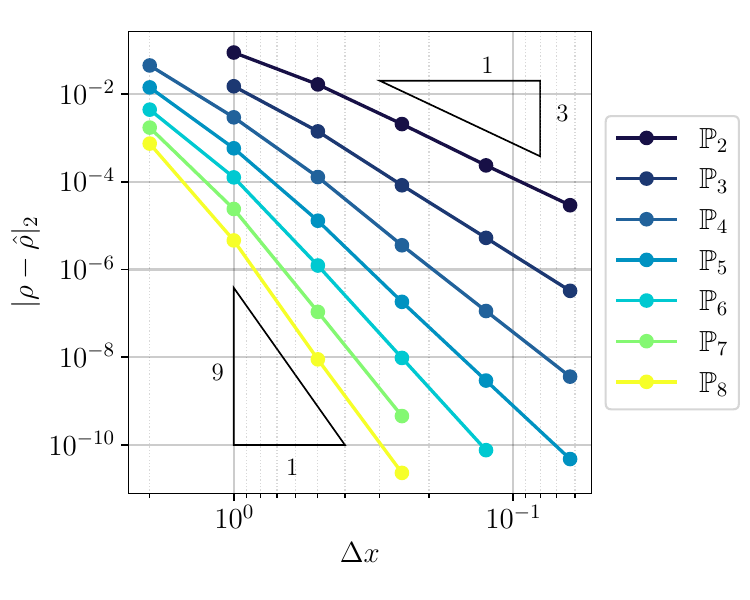} 
	\caption{Convergence of energy-stable split form DG scheme on LGL nodes using $p\in[2,8]$ for the manufactured solution.}
	\label{fig:h-convergence}
\end{figure}

The 2D Euler equation is solved in a square $[-1,1]^2$ with periodic boundary condition. The h-mesh resolution varies from $1^2$ to $32^2$ and the polynomial order is increased from 2 to 8. RK3 time integrator and local Laxfriedrichs riemann solver without stabilization terms are used. The computation is advanced in time up to $t = 1$ and the timestep is chosen sufficiently small to not influence the overall discretization error. The convergence test is carried out with the split-flux formulation on LGL interpolation points. The results in Fig.~\ref{fig:h-convergence} demonstrate that the expected design order is reached for all investigated cases, which verifies the correct implementation of the schemes.

\section{Analysis on the correction}
\label{sec: AppendixE}

\begin{figure}[h]
    \centering

    \begin{subfigure}[b]{1.0\textwidth}  
        \centering
        \adjustbox{valign=t}{\includegraphics[width=\textwidth, trim={0cm 0cm 0cm 0cm}, clip]{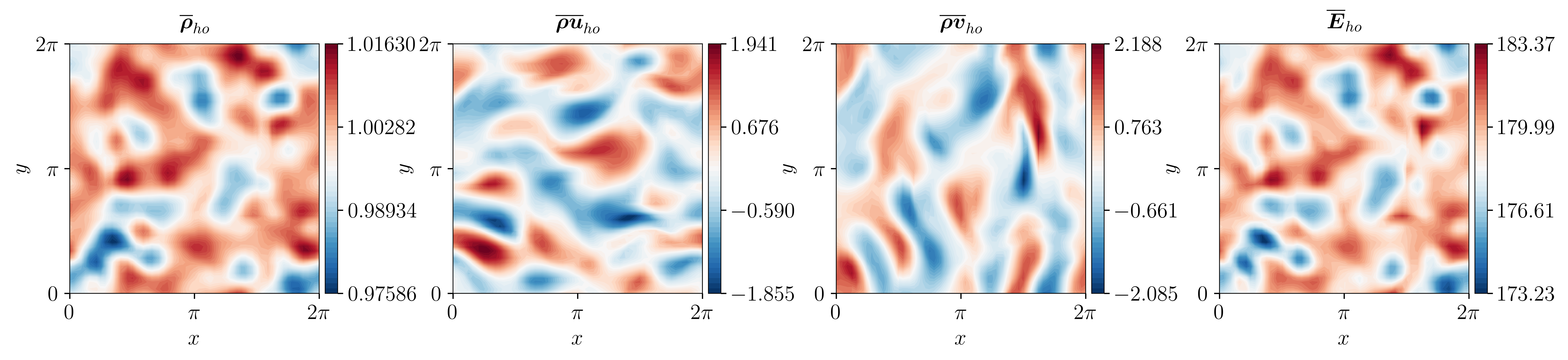}}  
        \caption{Flow fields of $\overline{\bm{q}}_{ho}$ at $t=1$. }
        \label{fig: Q_t=1}
    \end{subfigure}
    \begin{subfigure}[b]{1.0\textwidth}
        \centering
        \adjustbox{valign=t}{\includegraphics[width=\textwidth, trim={0cm 0cm 0cm 0cm}, clip]{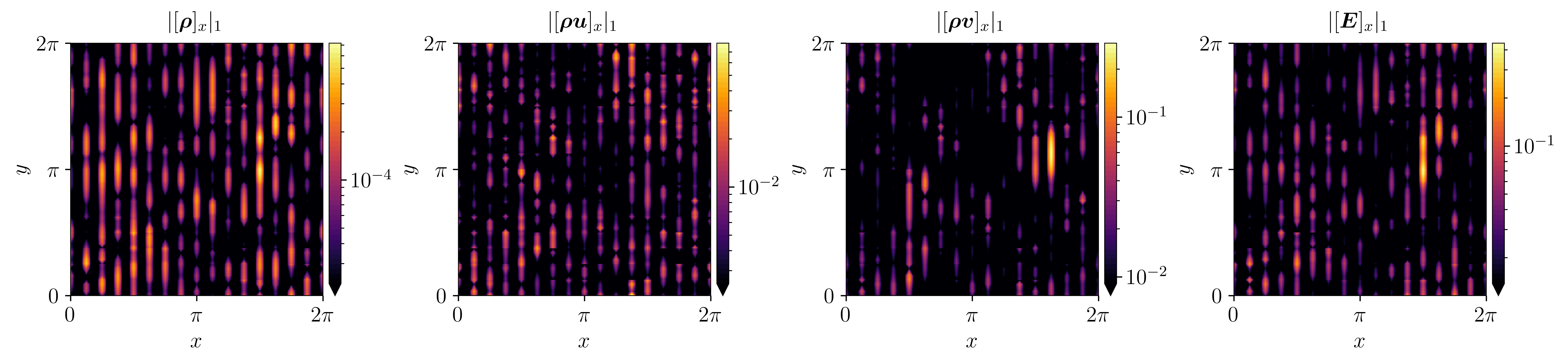}}  
        \caption{Jumps on the interfaces of $\overline{\bm{q}}_{ho}$ across the $x-$direction.}
        \label{fig: jumps_x_Q_t=1}
    \end{subfigure}

    \begin{subfigure}[c]{1.0\textwidth}
        \centering
        \adjustbox{valign=t}{\includegraphics[width=\textwidth, trim={0cm 0cm 0cm 0cm}, clip]{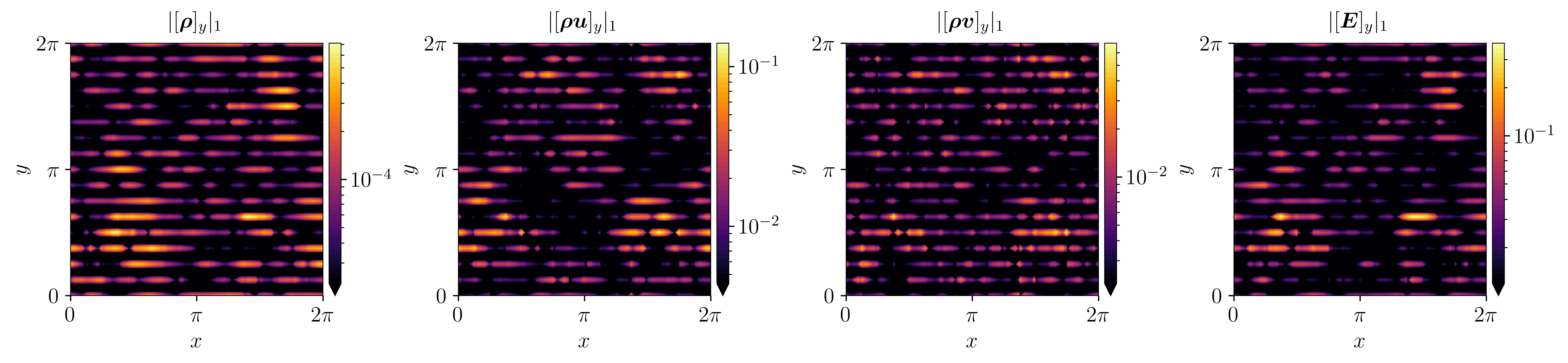}}  
        \caption{Jumps on the interfaces of $\overline{\bm{q}}_{ho}$ across the $y-$direction.}
        \label{fig: jumps_y_Q_t=1}
    \end{subfigure}

    \begin{subfigure}[d]{1.0\textwidth}
        \centering
        \adjustbox{valign=t}{\includegraphics[width=\textwidth, trim={0cm 0cm 0cm 0cm}, clip]{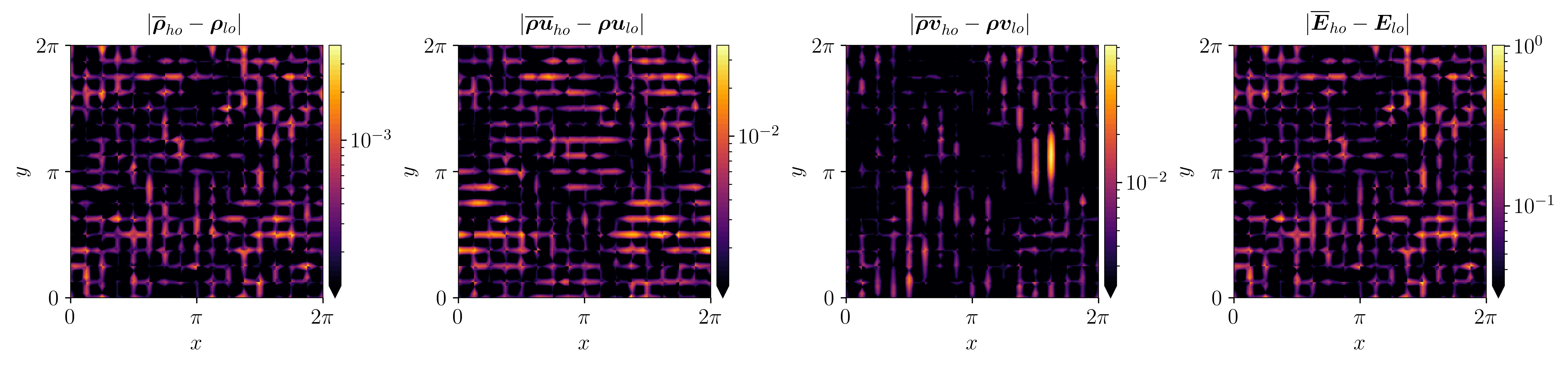}}  
        \caption{The target correction to learned at $t=1$ ($\overline{\bm{q}}_{ho}^1 - \bm{q}_{lo}^1$)}
        \label{fig: correction_Q_t=1}
    \end{subfigure}

    \caption{The contours of flow field, jumps across $x-$ and $y-$ direction and the correction.}
    \label{fig: flow_jumps_correction}
\end{figure}

Here we take the solution at $t=1$ for example to show the close relation between the jumps on the interfaces with the corresponding correction. The polynomial for low-order and high-order simulations are $p_{lo}=3$ and $p_{ho}=7$ respectively. Firstly, the flow fields of four conservative variables and their jumps on the interfaces are shown in Fig.~\ref{fig: Q_t=1}, Fig.~\ref{fig: jumps_x_Q_t=1} and Fig.~\ref{fig: jumps_y_Q_t=1}, where $[\;\cdot \;]_x$ and $[\;\cdot \;]_y$  denote the jumps across the interfaces normal to $x-$ and $y-$axis respectively. It can be seen that the jumps of $\bm{\rho u}$ and $\bm{\rho v}$ have obvious directional features: $[\bm{\rho u}]_y$ is larger than $[\bm{\rho u}]_x$ and the situation for $\bm{\rho v}$ is reversed. Therefore, we pay more attention to the distribution of $[\bm{\rho u}]_y$ and $[\bm{\rho v}]_x$. By comparing the highlights of $[\bm{\rho u}]_y$ and $[\bm{\rho v}]_x$ with the corrections for $\bm{\rho u}$ and $\bm{\rho v}$, we can easily find that they are highly related.

\end{document}